\documentclass[aps,prd,twocolumn,superscriptaddress,floatfix,nofootinbib,natbib,longbibliography]{revtex4-2}
\usepackage{graphicx,longtable,mathrsfs,color,array}
\usepackage[usenames,dvipsnames]{xcolor} 
\usepackage{amssymb,amsmath,mathtools,mathrsfs,slashed} 
\usepackage{braket}
\usepackage{epsfig,subfigure,placeins,float} 
\usepackage{booktabs,longtable,ctable,multirow} 
\usepackage{exscale,relsize} 
\usepackage[normalem]{ulem} 
\usepackage{enumerate}
\usepackage{enumitem}
\usepackage{dsfont}
\usepackage{ulem}
\usepackage{color}
\usepackage[qm]{qcircuit}
\usepackage{comment}
\usepackage[pdftex,
            pdftitle={Phase structure of the three-flavor Schwinger model with quantum computing},
            pdfauthor={Lena Funcke, Tobias Hartung, Karl Jansen, Stefan Kuehn, Marc-Oliver Pleinert, Stephan Schuster, Joachim von Zanthier},
            bookmarks,
            colorlinks,
            linkcolor=myblue,
            citecolor=mymagenta,
            menucolor=black,
            urlcolor=myblue,
            plainpages=false,
            pdfpagelabels,
            hypertexnames=false]{hyperref}

\definecolor{mymagenta}{RGB}{200, 0, 100}
\definecolor{myblue}{RGB}{45, 48, 146}
\definecolor{mygreen}{RGB}{0, 126, 0}
\definecolor{myorange}{RGB}{255, 136, 19}

\renewcommand{\sp}{\ensuremath{\sigma^+}}
\newcommand{\sm}{\ensuremath{\sigma^-}}

\renewcommand\vec{\boldsymbol}

\newcommand*{\QOQI}{Quantum Optics and Quantum Information Group,\\ Friedrich-Alexander-Universität Erlangen-Nürnberg, Staudtstr. 1, 91058 Erlangen, Germany}

\begin{document}
\title{Studying the phase diagram of the three-flavor Schwinger model in the presence of a chemical potential with measurement- and gate-based quantum computing}

\author{Stephan Schuster}
\affiliation{\QOQI}

\author{Stefan Kühn}
\affiliation{CQTA, Deutsches Elektronen-Synchrotron DESY, Platanenallee 6, 15738 Zeuthen, Germany}

\author{Lena Funcke}
\affiliation{Transdisciplinary Research Area ``Building Blocks of Matter and Fundamental Interactions'' (TRA Matter), University of Bonn, Bonn, Germany}
\affiliation{Helmholtz Institute for Radiation and Nuclear Physics (HISKP), University of Bonn, Nussallee 14-16, 53115 Bonn, Germany}
\affiliation{Bethe Center for Theoretical Physics (BCTP), University of Bonn, Nussallee 12, 53115 Bonn, Germany}

\author{Tobias Hartung}
\affiliation{Northeastern University - London, Devon House, St Katharine Docks, London, E1W 1LP, United Kingdom}
\affiliation{%
Khoury College of Computer Sciences, Northeastern University, 440 Huntington Avenue, 202 West Village H
Boston, MA 02115, USA
}

\author{Marc-Oliver Pleinert}
\author{Joachim von Zanthier}
\affiliation{\QOQI}

\author{Karl Jansen}
\affiliation{CQTA, Deutsches Elektronen-Synchrotron DESY, Platanenallee 6, 15738 Zeuthen, Germany}
\affiliation{Computation-Based Science and Technology Research Center, The Cyprus Institute, 20 Kavafi Street, 2121 Nicosia, Cyprus}

\date{\today}

\begin{abstract}
We propose an ansatz quantum circuit for the variational quantum eigensolver (VQE), suitable for exploring the phase structure of the multi-flavor Schwinger model in the presence of a chemical potential. Our ansatz is capable of incorporating relevant model symmetries via constrains on the parameters, and can be implemented on circuit-based as well as measurement-based quantum devices. We show via classical simulation of the VQE that our ansatz is able to capture the phase structure of the model, and can approximate the ground state to a high level of accuracy. Moreover, we perform proof-of-principle simulations on superconducting, gate-based quantum hardware. Our results show that our approach is suitable for current gate-based quantum devices, and can be readily implemented on measurement-based quantum devices once available.
\end{abstract}

\maketitle

\section{Introduction}
In recent years, methods originating from quantum information theory have emerged as a promising alternative for numerically investigating lattice field theories~\cite{Banuls2018a,Banuls2019,Banuls2020,Funcke2023a,DiMeglio:2023nsa}. In particular, methods based on tensor network states, a family of entanglement-based ansätze for the wave function of a quantum many-body system, have demonstrated their potential for overcoming the limitations of conventional Monte Carlo (MC) methods~\cite{Banuls2018a,Banuls2019}. Here, successful computations have been performed for lattice field theories in the presence of a topological term~\cite{Byrnes2002,Buyens2017,Funcke2019,Angelides:2023bme,Funcke2023} or chemical potentials~\cite{Banuls2016a, Silvi2016, Magnifico2021}, regimes in which conventional MC approaches suffer from the sign problem.

Moreover, quantum technologies heavily improved during recent years. This might provide another option for computationally investigating lattice field theories, even in potentially highly-entangled regimes such as out-of-equilibrium dynamics where tensor networks have limited applicability~\cite{Trotzky2012}. Using the Hamiltonian lattice formulation allows for directly simulating the theory under consideration on a quantum device, thus bypassing limitations of classical numerical methods. Several proof-of-principle experiments have demonstrated that such parameter regimes can indeed be investigated via simulations on quantum computers~\cite{Banuls2020,Funcke2023a,DiMeglio:2023nsa}.

In order to fully utilize the potential of currently available noisy intermediate-scale quantum (NISQ) devices, appropriate algorithms in combination with circuit optimization and error mitigation techniques are required~\cite{Lorenza1998,Endo2018,Funcke2020,Funcke2020a,Funcke2021a,Cai2020,Giurgica-Tiron2020,Berg2023}. The VQE has proven itself to be particularly suited for NISQ devices~\cite{Peruzzo2014,McClean2016}. This hybrid quantum-classical algorithm tries to approximate the ground state of a given Hamiltonian using a parametric quantum circuit as a variational ansatz, whose parameters are optimized in order to minimze the energy expectation value of the Hamiltonian. In order to achieve a good performance of the VQE, it is essential to choose a suitable ansatz. For an implementation on current NISQ devices, the variational ansatz must be sufficiently simple while still being expressive enough to reach the target ground state in the Hilbertspace. Additionally, it is desirable to incorporate the relevant symmetries of the underlying Hamiltonian in the variational ansatz to reduce the number of parameters in the classical optimization of the VQE. Very recently, it was shown that combining all these techniques carefully, a VQE for a (1+1)-dimensional gauge theory could be realized on hardware with up to 100 qubits~\cite{Farrell2023}.

In this work, we present an ansatz which is suitable for VQE of the multi-flavor Schwinger model in the presence of a chemical potential. This model is, in general, inaccessible with MC methods in the presence of a chemical potential due to the sign problem. Focusing on the case of three fermion flavors, we develop a suitable variational ansatz circuit for the model. We show via classical simulations of the VQE that the ansatz is able to capture the relevant physics at a low circuit depth. Additionally, we demonstrate how to incorporate symmetries of the model in our ansatz, reducing the number of variational parameters in the classical optimization. Besides presenting our ansatz in the circuit model, we demonstrate that it also lends it to the formalism of measurement-based quantum computation~\cite{Raussendorf2001,Raussendorf2003} allowing for a future implementation on a one-way quantum computer. Finally, we show proof-of-principle results on IBM's circuit-based superconducting quantum hardware.

The paper is organized as follows. In Sec.~\ref{sec:model}, we briefly introduce the staggered Hamiltonian lattice formulation for the multi-flavor Schwinger model. Moreover, we discuss its symmetries and the phase structure in the presence of a chemical potential. We proceed with presenting our ansatz for the VQE in Sec.~\ref{sec:ansatz}. Finally, we show the performance results of the ansatz in various parameter regimes in Sec.~\ref{sec:results}, before concluding in Sec.~\ref{sec:conclusion}.

\section{Model and Methods}
\label{sec:model}
\subsection{Lattice Hamiltonian and spin formulation}
\label{sec:hamiltonian}
For our study, we use a Hamiltonian lattice formulation of the Schwinger model with staggered fermions, which reads~\cite{Kogut1975,Hamer1997,Banuls2016a} 
\begin{align}
    \begin{aligned}
         H = &-\frac{i}{2a}\sum_{n=0}^{N-2}\sum_{f=0}^{F-1}\left(\phi_{n,f}e^{i\theta_n}\phi_{n+1,f}-\mathrm{h.c.}\right)\\
             &+\sum_{n=0}^{N-1}\sum_{f=0}^{F-1}\left(m_f(-1)^n +\kappa_f \right)\phi_{n,f}\phi_{n,f}\\
             &+ \frac{g^2a}{2}\sum_{n=0}^{N-2} L_n^2,
    \end{aligned}
    \label{eq:hamiltonian}
\end{align}
for $F$ fermion flavors on a lattice with $N$ sites and spacing $a$. In the expression above, $\phi_{n,f}$ ($\phi_{n,f}^\dagger$) are single-component fermionic fields annihilating (creating) a particle of flavor $f$ at site $n$. The parameter $g$ denotes the coupling, while $m_f$ and $\kappa_f$ correspond to the bare mass and the bare chemical potential for flavor $f$. The operators $L_n$ and $e^{i\theta_n}$ act on the links in between two adjacent matter sites $n$ and $n+1$, where $L_n$ and $\theta_n$ are canonical conjugate variables fulfilling the commutation relation $[\theta_n,L_{n'}] = i\delta_{nn'}$. Thus, in the eigenbasis of the electric field operator $L_n$, $e^{i\theta_n}$ corresponds to the lowering operator. The first line of the Hamiltonian in Eq.~\eqref{eq:hamiltonian} represents the kinetic term, corresponding to fermionic hopping while simultaneously chaning the electric field. The first summand in the second line corresponds to the mass and the chemical potential term whereas the second one represents the electric energy. The physical states $\ket{\psi}$ of the Hamiltonian in Eq.~\eqref{eq:hamiltonian} have to fulfill Gauss law, i.e., they have to be eigenstates of the operators
\begin{align}
    G_n = L_n  - L_{n-1} - Q_n.
    \label{eq:gauss_law}
\end{align}
The operators $G_n$ are the generators for time-independent gauge transformations, and $Q_n = \sum_{f=0}^{F-1}\phi_{n,f}^\dagger\phi_{n,f}-\frac{F}{2}(1-(-1)^n)$ is the staggered charge. The integer eigenvalues $q_n$ of $G_n$ correspond to static external charges. For the rest of the paper, we choose to work in the sector of vanishing external charges, $G_n\ket{\psi} = 0$, for each site $n$.

For open boundary conditions, Eq.~\eqref{eq:gauss_law} allows us to reconstruct the electric field values purely from the fermionic charge content of the sites after fixing the value $l_{-1}$ of the electric field on the left boundary, $L_n = l_{-1}+ \sum_{k=0}^n Q_k$. Inserting this into Eq.~\eqref{eq:hamiltonian}, applying a residual gauge transformation, and making the resulting Hamiltonian dimensionless~\cite{Hamer1997,Sala2018,Angelides:2023bme}, we find
\begin{align}
    \begin{aligned}
         W =& -ix\sum_{n=0}^{N-2}\sum_{f=0}^{F-1}\left(\phi^\dagger_{n,f}\phi_{n+1,f}-\mathrm{h.c.}\right)\\
         &+\sum_{n=0}^{N-1}\sum_{f=0}^{F-1}\left(\mu_f(-1)^n +\nu_f \right)\phi^\dagger_{n,f}\phi_{n,f} \\
         &+ \sum_{n=0}^{N-2} \left( \sum_{k=0}^n Q_k +  l_{-1} \right)^2.
    \end{aligned}
    \label{eq:hamiltonian_dimensionless}
\end{align}
In the expression above, we use the dimensionless quantities $x \equiv 1/(ag)^2$, $\mu_f \equiv 2\sqrt{x}m_f/g$, and $\nu_f \equiv 2\sqrt{x}\kappa_f/g$. The parameter $l_{-1}$ represents a constant background field, and corresponds to the lattice discretization of a topological $\theta$-term~\cite{Coleman:1976uz,Byrnes2002,Funcke2019}. Here, we focus on the case of a vanishing background field and set $l_{-1} =0$ for the rest of the paper. Note that in Eq.~\eqref{eq:hamiltonian_dimensionless} the gauge fields are no longer present. We therefore  obtain a formulation directly restricted to the gauge invariant subspace of the theory. This comes at the expense of creating long-range interactions between the fermionic degrees of freedom.

In this work, we aim at studying the phase structure of the model using VQE. In order to measure the expectation value of the dimensionless Hamiltonian $W$ on a quantum device, we choose to translate the fermionic degrees of freedom into spins using a Jordan-Wigner transformation~\cite{Hamer1997,Banuls2016a}. The fermionic operators appearing in the kinetic term, the mass term and the electric energy term of the Hamiltonian are mapped according to
\begin{align}
    \phi^\dagger_{n,f}\phi_{n+1,f} &\to \sp_{n,f}(iZ_{n,f})\dots(iZ_{n+1,f-1}) \sm_{n+1,f},\\
    \phi^\dagger_{n,f}\phi_{n,f} &\to \frac{1}{2}\left(Z_{n,f} + \mathds{1}\right).
\end{align}
In the expression above, $\sigma^\pm \equiv (X \pm i Y)/2$ and $X$, $Y$ and $Z$ are the usual Pauli matrices. Inserting this transformation into Eq.~\eqref{eq:hamiltonian_dimensionless}, we obtain a spin chain of length $NF$. This allows for obtaining the expectation value of the Hamiltonian by arranging the different Pauli terms into commuting groups and measuring them individually.

Regarding the symmetries of our model Hamiltonian, note that $W$ conserves the total charge $Q_\mathrm{tot} = \sum_{n=0}^{N-1}Q_n$. We will therefore restrict ourselves to the sector of vanishing total charge, $Q_\mathrm{tot}\ket{\psi} = 0$. Additionally, for an odd number of flavors $F$ and the special choice of $\nu_f=-\nu_{F-1-f}$, $\mu_f=\mu_{F-1-f}$, $f\leq \lfloor N/2\rfloor$, the spin Hamiltonian is invariant under flipping all spins, followed by a spatial reflection around the lattice center. The corresponding unitary $S$ implementing the transformation acts on the Pauli matrices occurring in the Hamiltonian as
\begin{equation}
    \label{eq:SymmOp}
    \mathnormal{S}A_j\mathnormal{S}^\dagger = (XAX)_{NF-1-j},
\end{equation}
where $A\in\{X,Y,Z,\sigma^\pm\}$. Note here that the index of the spin operators does not correspond to the fermionic sites, since each of the fermion flavors within a site is mapped onto a spin degree of freedom by the Jordan-Wigner transformation.

\subsection{Mass renormalization for staggered fermions\label{sec:MassShift}}

In general, the bare fermion mass that is chosen as a parameter in the lattice discretization of a field theory does not correspond to the physical fermion mass, and one has to consider renormalization effects to determine the physical fermion mass. Recently, an analytical prediction for the additive mass renormalization of staggered fermions was derived for the lattice Schwinger model with periodic boundary conditions, both for the single-flavor~\cite{Dempsey:2022nys} and the multi-flavor~\cite{Dempsey:2023gib} case. This derivation was based on enforcing a discrete spurious chiral symmetry given by a translation of one lattice site followed by shifting $\theta$ by $\pi$. The resulting mass shift (MS) in units of the coupling is given by~\cite{Dempsey:2022nys, Dempsey:2023gib}
\begin{align}
   \frac{\textrm{MS}}{g}= \frac{F}{8\sqrt{x}}.
   \label{eq:renormalized_mass_staggered}
\end{align}
Thus, the renormalized fermion mass $m_r$ is not equal to the bare lattice fermion mass $m_f$ defined in the original lattice Hamiltonian~\eqref{eq:hamiltonian}, but receives an additive mass renormalization, such that $(m_r/g) = (m_f/g) + ({\rm MS}/g)$.

More generally, the additive mass renormalization for the Schwinger model can be determined numerically using the method proposed in Refs.~\cite{Angelides:2022pah,Angelides:2023bme}, i.e., by identifying the point $m_r/g=0$ at which the electric field density vanishes, independently of the boundary conditions. This approach was demonstrated for the single-flavor Schwinger model in Refs.~\cite{Angelides:2022pah,Angelides:2023bme}, where it was found that the method can significantly improve the convergence towards the continuum limit. Moreover, for sufficiently large volumes, $V=N/\sqrt{x}\gtrsim 30$, the results for the mass shift obtained with open boundary conditions were shown to agree with the theoretical prediction for periodic boundary conditions~\cite{Dempsey:2022nys}. Note that this technique can be straightforwardly extended to the multi-flavor Schwinger model, because the condition that the electric field density vanishes at the point $m_r/g=0$ holds true also for multiple fermion flavors~\cite{Coleman:1976uz}. In particular, the method can be readily implemented in a VQE setting by measuring the electric field density in the ground state as a function of the lattice mass $m_f/g$ and by determining the point at which the electric field density vanishes.

\subsection{Phase structure of the model}
\label{sec:phasestructure}
For the massless case, the phase structure of the continuum model in a finite volume with periodic boundary conditions has been determined analytically in Refs.~\cite{Narayanan2012,Lohmayer2013}. It was found that the model undergoes an infinite number of first-order phase transitions, whose locations only depend on the difference of the chemical potentials $\nu_f - \nu_{f'}$ with respect to a single, arbitrarily chosen flavor $f'$. Each of the phases is characterized by the particle numbers
\begin{equation}
    N_f=\sum_{n=0}^{N-1}\phi^\dagger_{n,f}\phi_{n,f}
\end{equation}
of the different flavors $f$. For two flavors of fermions, Ref.~\cite{Banuls2016a} showed that the picture also qualitatively persists on a finite lattice with open boundary conditions. Moreover, when taking the continuum limit, the results obtained from the lattice calculations agreed with the theoretical prediction of Ref.~\cite{Narayanan2012}. 
Additionally, it was demonstrated for the case of two fermion flavors that these first-order phase transitions also occur for nonvanishing bare fermion mass~\cite{Banuls2016a}, and also for quantum electrodynamics in 2+1 dimensions~\cite{Bender2023}. Note that the phase structure of the model can in general not be assessed numerically with conventional MC methods, as these suffer from a sign problem as soon as $\sum_{f}\kappa_f\neq 0$.

A theoretical prediction for the locations of the phase transitions in the lattice model can be obtained following the ideas in Ref.~\cite{Banuls2016a}. Since the Hamiltonian in Eq.~\eqref{eq:hamiltonian_dimensionless} commutes with $N_f$ for each flavor, the energy eigenstates are simultaneously eigenstates of the particle number operators. Consequently, $W$ has a block-diagonal form in which each block can be labelled with $(N_0,\cdots,N_{F-1})$. In order to derive a theoretical prediction for the phase transition points it will be convenient to rewrite $W$ as
\begin{equation}
    W = \sum_{f=0}^{F-1}\nu_f N_f + W_\text{aux},
\end{equation}
where $W_\text{aux}$ contains all parts of $W$ which are independent of $\nu_f$~\cite{Banuls2016a}. The minimum eigenvalue in a given block is then
\begin{equation}
    \begin{split}
    \label{eq:energyPT}
    &E_{(N_0,\cdots,N_{F-1})}(\nu_0,\cdots,\nu_{F-1}) = \\
    &=\sum_{f=0}^{F-1}\nu_f N_f + E_\text{min}(W_\text{aux}\lvert_{(N_0,\cdots,N_{F-1})}),
    \end{split}
\end{equation}
where $E_\text{min}(W_\text{aux}\lvert_{(N_0,\cdots,N_{F-1})})$ is a block-dependent constant. Therefore, measuring the ground-state energy $E_{(N_0,\cdots,N_{F-1})}(\nu_0,\cdots,\nu_{F-1})$ as well as all particle numbers $N_f$ at one point $(\nu_0,\cdots,\nu_{F-1})$ is sufficient to determine this constant for the whole block. We can now simplify Eq.~\eqref{eq:energyPT} to the case of three fermion flavors, $F=3$, considered throughout the rest of the paper and set $\nu_2 = -\nu_0$\footnote{This restriction is contained in the considered symmetry subspace as introduced in Sect.~\ref{sec:hamiltonian}, see Eq.~\eqref{eq:SymmOp}, but it is also used in our simulations outside of this symmetry subspace for simplicity (cf. Sec.~\ref{sec:results_sign_prob}).}, which yields
\begin{equation}
\begin{split}
\label{eq:energyPT3F}
    E_{(N_0,N_1,N_2)}(\nu_0,\nu_1) = &\nu_0 (N_0-N_2) + \nu_1 N_1 \\
    &+ E_\text{min}(W_\text{aux}\lvert_{(N_0,N_1, N_2)}).
\end{split}    
\end{equation}
In the following, we will consider $\nu_0$ as a variable to scan through the phase diagram and $\nu_1$ to be a constant. This allows us to derive an analytical expression for the transition points $\nu_0\lvert_\text{jump}$.
During a first-order phase transition, it becomes energetically favorable to go from one block characterized by $(N_0,N_1,N_2)$ to a neighbouring block with $(\bar{N}_0,\bar{N}_1,\bar{N}_2)$. Directly at the critical point, the energy levels of the neighbouring blocks are degenerate, $E_{(N_0, N_1,N_2)}(\nu_0,\nu_1)=E_{(\bar{N}_0,\bar{N}_1,\bar{N}_2)}(\nu_0,\nu_1)$. Using this equality together with Eq.~\eqref{eq:energyPT3F}, we obtain the following expression for the critical point:
\begin{equation*}
\begin{split}
    \nu_0\lvert_\text{jump}=&\frac{E_\text{min}(W_\text{aux}\lvert_{(\bar{N}_0,\bar{N}_1, \bar{N}_2)})-E_\text{min}(W_\text{aux}\lvert_{(N_0,N_1, N_2)})}{(N_0-\bar{N}_0)-(N_2-\bar{N}_2)}\\
    & - \frac{\nu_1(N_1-\bar{N}_1)}{(N_0-\bar{N}_0)-(N_2-\bar{N}_2)}.
    \end{split}
\end{equation*}
As outlined above, the block-dependent constants $E_\text{min}(W_\text{aux}\lvert_{(N_0,N_1, N_2)})$ can be obtained from measuring $E_{(N_0,N_1,N_2)}$ and $N_f$ at an arbitrary point $(\nu_0^*, \nu_1^*)$ inside each block. In particular, the point can be far away from the transition point. Considering this in the above equation, we find for the transition point
\begin{equation}
\begin{split}
    \nu_0\lvert_\text{jump}=&\frac{E_{(\bar{N}_0,\bar{N}_1,\bar{N}_2)}(\bar{\nu}^*_0,\nu_1)-\bar{\nu}^*_0(\bar{N}_0-\bar{N}_2)}{(N_0-\bar{N}_0)-(N_2-\bar{N}_2)}\\
    & - \frac{E_{(N_0, N_1,N_2)}(\nu^*_0,\nu_1)-\nu^*_0(N_0-N_2)}{(N_0-\bar{N}_0)-(N_2-\bar{N}_2)}.
    \end{split}
    \label{eq:nujump_3F}
\end{equation}
Alternatively, we can label each block of $W$ with the differences in the particle numbers $\Delta N_f = N_f - N_{f'}$ with respect to a single, arbitrarily chosen flavor $f'$ and the total particle number $\sum_f N_f$. Equation~\eqref{eq:nujump_3F} can thus be rewritten in terms of the particle number differences $\Delta N_f = N_f - N_1$ for $f'=1$, yielding
\begin{equation}
\begin{split}
    \nu_0\lvert_\text{jump}=&\frac{E_{(\Delta\bar{N}_0,\Delta\bar{N}_1,\Delta\bar{N}_2)}(\bar{\nu}^*_0,\nu_1)-\bar{\nu}^*_0(\Delta\bar{N}_0-\Delta\bar{N}_2)}{(\Delta N_0-\Delta\bar{N}_0)-(\Delta N_2-\Delta\bar{N}_2)}\\
    & - \frac{E_{(\Delta N_0, \Delta N_1, \Delta N_2)}(\nu^*_0,\nu_1)-\nu^*_0(\Delta N_0-\Delta N_2)}{(\Delta N_0-\Delta \bar{N}_0)-(\Delta N_2-\Delta \bar{N}_2)}.
    \end{split}
    \label{eq:nujump_3F_dN}
\end{equation}
We will use Eq.~\eqref{eq:nujump_3F_dN} in Sec.~\ref{sec:results_ibm} to predict the relevant phase transition points from our hardware data.

\section{VQE protocol}
\label{sec:ansatz}
In the following, we will introduce our parametric ansatz circuit for the VQE protocol. After introducing the ansatz, we will show how to incorporate the symmetries of the model into the ansatz, which allows for reducing the number of variational parameters. We will give a general description of the ansatz, which can be straightforwardly realized on any circuit-based quantum device, in Sect.~\ref{sec:QCM}. Additionally, we show that our ansatz lends itself also to the one-way model of quantum computation in Sect.~\ref{sec:1wqc}, which opens up the possibility of implementing our algorithm on measurement-based quantum computers in the future.

\subsection{Quantum circuit model}
\label{sec:QCM}
\begin{figure}[htp!]
    \centering
    \includegraphics[width=1.0\columnwidth]{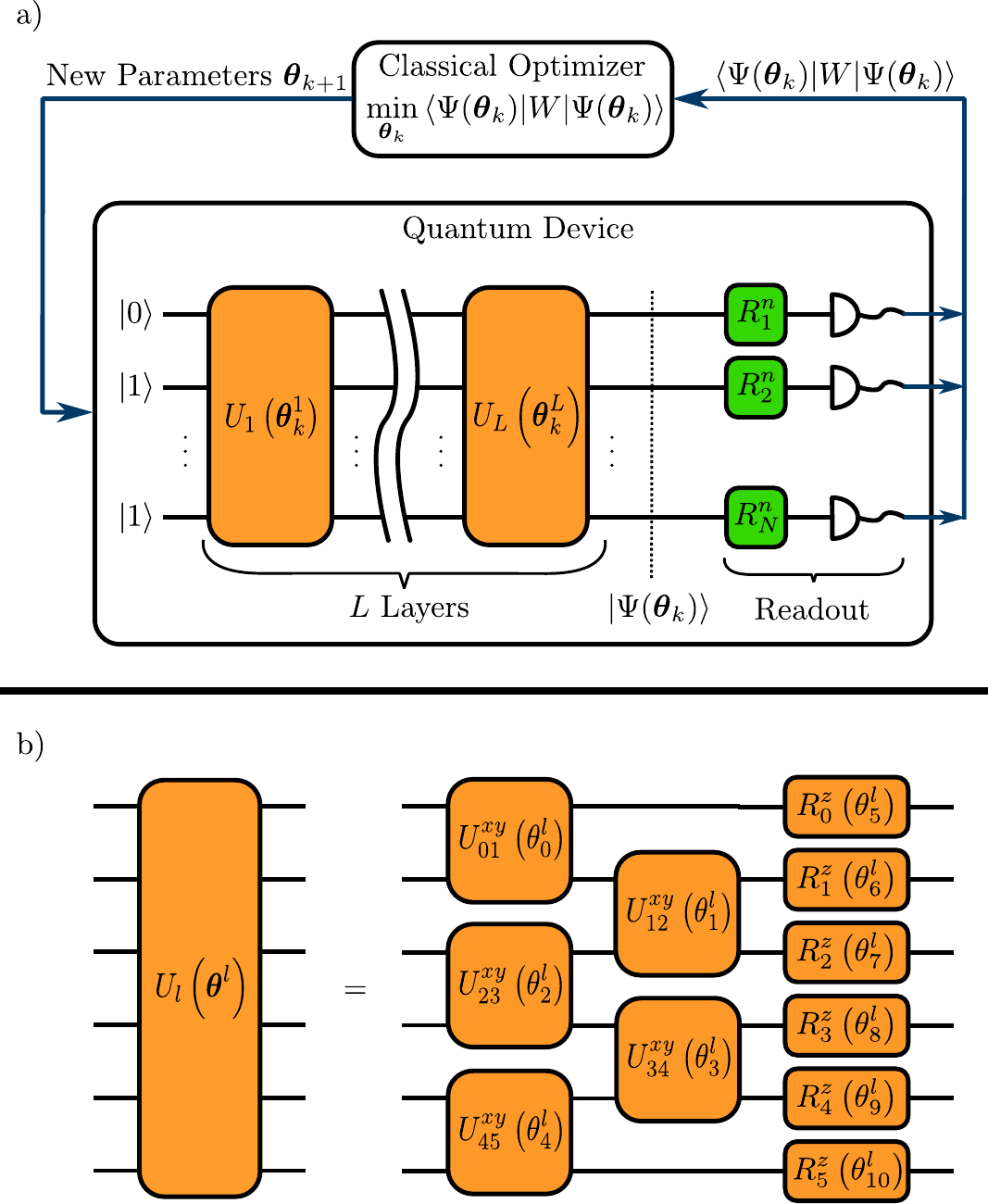}
    \caption{a) VQE protocol using a layered circuit ansatz $U_l(\vec{\theta}^l_k)$ with $L$ layers to prepare trial states $\{\ket{\Psi(\vec{\theta}_k)}\}$ from initial state $\ket{\Psi_0}=\ket{0101\cdots1}$ and a classical optimizer to iteratively find the approximate ground state of $W$. Here, $R^n_i$ indicate post rotations needed to measure the energy expectation value $\braket{\Psi(\vec{\theta}_k)|W|\Psi(\vec{\theta}_k)}$. b) Illustration of the circuit corresponding to Eq.~\eqref{eq:QuAlg} for six qubits representing three lattice sites.}
    \label{fig:vqe_protocol_decomp}
\end{figure}

The VQE utilizes a parametric circuit ansatz realized on a quantum device, in conjunction with a closed feedback-loop with a classical optimization routine find an approximation for the ground state of our model Hamiltonian $W$~\cite{Peruzzo2014}. The quantum device is capable of efficiently preparing a set of trial states $\{\ket{\Psi(\vec{\theta}_k)}\}$ depending on the variational parameters $\vec{\theta}_k \in \mathbb{R}^p$, starting from a fixed initial state $\ket{\Psi_0}$. After the state preparation, the energy $\braket{\Psi(\vec{\theta}_k)|W|\Psi(\vec{\theta}_k)}$ is measured and handed to the classical optimizer as a cost function. The optimizer attempts to minimize this cost function by iteratively adjusting the algorithm parameters, $\vec{\theta}_k \to \vec{\theta}_{k+1}$, and handing them back to the quantum device (cf.\ Fig.\ref{fig:vqe_protocol_decomp}a)).

The ansatz we propose, consists of a series of $L$ layers. Each layer $l$ realizes the following unitary operation on the $NF$ qubits corresponding to a system with $N$ sites
\begin{equation}
\label{eq:QuAlg}
\begin{split}
    U_l(\vec{\theta}^l)= &\prod_{i=0}^{NF-1}R^z_i(\theta^l_{NF-1+i})\prod_{i \text{ odd}}U^{xy}_{i,i+1}(\theta^l_i)\\
    &\prod_{i \text{ even}}U^{xy}_{i,i+1}(\theta^l_i),
    \end{split}
\end{equation}
where
\begin{align}
    U_{ij}^{xy}(\theta)&=e^{-i\frac{\theta}{2}(X_iX_{j}+Y_iY_j)},\label{eq:xx-yy-gate}\\
    R_i^z(\theta)&= e^{-i\frac{\theta}{2}Z_i}. \label{eq:z-rotation_def}
\end{align}
and $\vec{\theta}^l$ is a real vector representing the parameters of layer $l$. One layer thus consists of two-qubit entangling gates arranged in an odd-even pattern, followed by local $R^z$-rotations on all qubits (cf.\ Fig.~\ref{fig:vqe_protocol_decomp}b) for an illustration). All $L$ layers are applied sequentially to the initial state $\ket{\Psi_0}$, realizing the trial state
\begin{equation*}
    \ket{\Psi(\vec{\theta})}=\prod_{l=1}^{L}U_l(\vec{\theta}^l)\ket{\Psi_0},
\end{equation*}
with\footnote{We consider $\vec{\theta}$ and all $\vec{\theta}^l$ to be row vectors.} $\vec{\theta}=\left(\vec{\theta}^1,\vec{\theta}^2,\cdots,\vec{\theta}^L\right)$ and $U_l(\vec{\theta}^l)$ given in Eq.~\eqref{eq:QuAlg}. 
The whole VQE protocol is shown schematically in Fig.~\ref{fig:vqe_protocol_decomp}a).
As one can easily show, the gate operations in Eq.~\eqref{eq:QuAlg} conserve the total charge. Moreover, we can ensure that our algorithm is invariant under the symmetry operation from Eq.~\eqref{eq:SymmOp} by restricting the parameters in Eq.~\eqref{eq:QuAlg} to
\begin{align}
    \theta^l_i&=\theta^l_{NF-2-i}\quad \text{for }i\in[0,NF-2]\label{eq:QuAlgSymmRest_oe}\\
        \theta^l_i&=-\theta^l_{3NF-3-i}\quad \text{for }i\in[NF-1,2NF-2]\label{eq:QuAlgSymmRest_z}
\end{align}
where Eq.~\eqref{eq:QuAlgSymmRest_oe} corresponds to the parameter restriction for the two qubit gates, $U^{xy}_{i,i+1},$ of $U_l$ and Eq.~\eqref{eq:QuAlgSymmRest_z} represents the parameter restriction for the local $z$-rotations, $R^z_i$, of $U_l$.
Equations~\eqref{eq:QuAlgSymmRest_oe} and \eqref{eq:QuAlgSymmRest_z} together reduce the number of optimization parameters from $p=2NF-1$ per layer to $p=NF$, when investigating this symmetry subspace of our model. For investigations outside of this subspace, i.e., $\kappa_f\neq-\kappa_{F-1-f}$ or $m_f\neq m_{F-1-f}$, $f\leq \lfloor F/2\rfloor$, we can simply relax the parameter restriction of Eqs.~\eqref{eq:QuAlgSymmRest_oe} and \eqref{eq:QuAlgSymmRest_z}, and treat all parameters as independent.\\
To obtain a state in the correct symmetry sector, not only the ansatz must respect the symmetry, but also the initial state $\ket{\Psi_0}$ must be in the same symmetry sector that one targets (cf.\ Sect.~\ref{sec:hamiltonian}). In our simulations, we will use the Neel state $\ket{\Psi_0}=\ket{0101\cdots1}$, which has vanishing total charge and remains invariant under the symmetry operation from Eq.~\eqref{eq:SymmOp}. This state can easily be obtained from the state $\ket{00\cdots0}$ the qubits are typically initialized in on circuit-based devices by flipping the corresponding qubits with X-gates.

\subsection{One-way model of quantum computation}
\label{sec:1wqc}
Besides the widespread circuit model of quantum computation, also measurement-based approaches exist. While it can be shown that both models are equivalent, such measurement-based models could provide a better suited path to quantum computation for certain physical platforms, e.g., photonic setups~\cite{Marqversen2023, Nielsen2004}. In particular, VQE protocols for measurement-based quantum hardware~\cite{Ferguson:2020qyf} as well as hybrid approaches~\cite{Chan2023} have already been discussed. Here, we will translate our parametric ansatz circuit into the one-way model of quantum computation, a measurement-based model proposed in Refs.~\cite{Raussendorf2001, Raussendorf2003}. This allows for implementing our algorithm on such one-way quantum computer hardware in the future. In the following, we introduce the basics of the one-way model of quantum computation before we translate our ansatz circuit to this framework.

\subsubsection{Basic concepts of the one-way model of quantum computation}
\begin{figure}[htb]
    \centering
    \includegraphics[width=1.0\columnwidth]{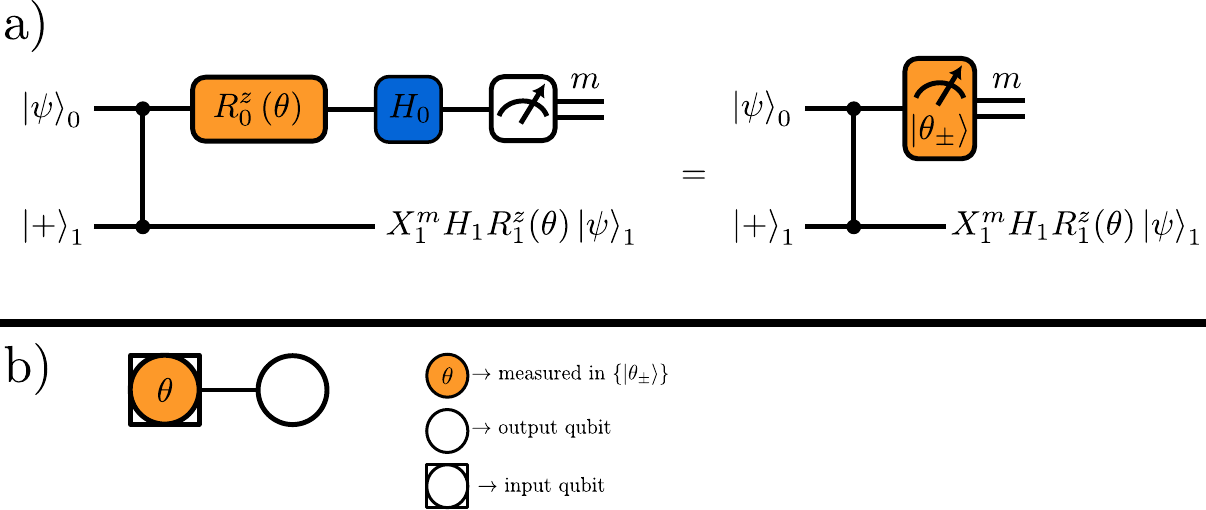}
    \caption{a) Illustration of the single-qubit teleportation protocol, the upper quantum wire represents qubit 0, the lower one qubit 1. The measurement in the computational basis following the $R^z$-rotation gate and the Hadamard gate in the left part effectively corresponds to a measurement in the rotated $X$-basis $\ket{\theta_\pm} \equiv \left(\ket{0}\pm e^{-i\theta}\ket{1}\right)/\sqrt{2}$ as shown in the right part. The measurement outcome $m$ takes binary values, $m\in\{0, 1\}$. b) The corresponding graph state for $\ket{\psi}_0=\ket{+}_0$ and measurement pattern for the one-way computation.}
    \label{fig:single_qubit_teleportation}
\end{figure}

The basic idea of the one-way model of quantum computation is to perform a series of single-qubit measurements on specific qubits of a highly entangled resource state, thereby effectively performing unitary operations on the unmeasured qubits. Typically a special class of entangled states called graph states are used a resource. As the name suggests, these can be conveniently represented as mathematical graph: the vertices of the graph correspond to qubits initialized in the state $\ket{+}= (\ket{0} + \ket{1})/\sqrt{2}$, the edges connecting two qubits correspond to a controlled-$Z$ gate applied between the qubits. For example, Fig.~\ref{fig:single_qubit_teleportation}b) shows the two-qubit linear graph state given by
\begin{equation}
    CZ_{01}\ket{+}_0\ket{+}_1,
\end{equation}
and Fig.~\ref{fig:Rz-gate_sqtp_mbqc}b) shows the three-qubit linear graph state given by
\begin{equation}CZ_{01}CZ_{12}\ket{+}_0\ket{+}_1\ket{+}_2.
\end{equation}
Here, $CZ_{ij}$ represents a controlled-$Z$ gate between qubits $i$ and $j$, and $\ket{+}_i$ corresponds to the state of qubit $i$.

Unitary operations can be implemented on the basis of the generalized single-qubit teleportation scheme, which is illustrated in Fig.~\ref{fig:single_qubit_teleportation}a).
This circuit allows for teleporting any single-qubit quantum state $\ket{\psi}$ from qubit $0$ to qubit $1$ while applying a $z$-rotation and a Hadamard gate to $\ket{\psi}$. Additionally, if the measurement outcome of qubit 1 is $m=1$, a Pauli-$X$ operator is added. The application of $R^z(\theta)H$ together with the measurement in the computational basis correspond to a measurement in the rotated $X$-basis, $\ket{\theta_\pm}=\left(\ket{0}\pm e^{-i\theta}\ket{1}\right)/\sqrt{2}$ (cf.\ right-hand side of Fig.~\ref{fig:single_qubit_teleportation}a)). The teleportation example shows that applying a measurement to one of the qubits effectively implements a set of unitary operations on the state of the qubit 0. This is because the two qubits are entangled at the beginning with a $CZ$ gate. After the measurement, qubit 0 is projected to one of the basis states and is no longer entangled with qubit 1. Moreover, since the state of qubit 0 is in general not one of the basis states $\ket{\theta_{\pm}}$, the output in qubit 1 is probabilistic and depends on the measurement outcome $m$ obtained for qubit 0, which is reflected in the $X^m$ in the final state.

The circuit for the teleportation scheme can also be illustrated conveniently in form of a mathematical graph, as shown in Fig.~\ref{fig:single_qubit_teleportation}b). Choosing $\ket{\psi}_0 = \ket{+}_0$ for the input qubit 0 of the computation, the initial entangling $CZ$-gate creates a graph state, represented by two vertices connected by an edge. Subsequently the input qubit is measured in the $\ket{\theta_\pm}$ basis, indicated by the color and label in the left vertex which represents qubit 0. The output state, i.e.,\ the final result of the circuit, is then contained in the vertex not carrying any color nor label, which represents the output qubit of this computation.

This principle can now be generalized to perform computational tasks involving a series of unitary operations. In particular, also two-qubit gates can be implemented with this method which requires two-dimensional graph states. To this end, a sequence of single-qubit measurements is applied to qubits of a graph state. These are either performed in the eigenbasis of one of the Pauli operators $X$, $Y$, $Z$ or in the rotated $X$-basis $\ket{\theta_\pm}$. Moreover, the measurement basis chosen at one point of the sequence can depend on the previous measurement outcome. Hence, in general the measurements have to be chosen adaptively~\cite{Nielsen2006, Kok2010}.  At the end of the sequence of measurements, the subset of qubits that have not been measured, called output qubits, contains the final state. Qubits that have been measured, have been projected to one of the basis states and are no longer entangled with the subset of qubits that have not been measured. As illustrated already in the teleportation example, the measurement outcome is generally probabilistic, resulting in a set of possible output states for the unmeasured output qubits. In order to make the computation deterministic, one needs to keep track of the outcome of each individual measurement and apply so-called Pauli corrections to the output qubits depending on the measurement outcome. The Pauli corrections can be shown to consist of Pauli $X$- and $Z$-operators. Moreover, it is possible to compensate for the Pauli corrections at the end of a computation instead of after each individual measurement. 

Following Refs.~\cite{Nielsen2006, Kok2010}, we can translate circuits into the one-way model by decomposing the operations into single-qubit $z$-rotations, $R^z$, Hadamard gates, $H$, and $CZ$-gates. These operations can then be further decomposed into single-qubit-teleportation-based schemes as illustrated above (cf.\ Fig.~\ref{fig:single_qubit_teleportation}). The required graph state as well as all (adaptive) measurement bases and the resulting Pauli corrections can be extracted from this form.

\subsubsection{Formulating the ansatz in the one-way model of quantum computation}

In the gate-based representation for our ansatz in Eq.\eqref{eq:QuAlg}, we have two basic types of unitaries, single-qubit $R^z$ rotation gates and $U^{xy}$ entangling gates. Here we show how to represent both of them in the one-way gate set $\{R^z,\, H,\, CZ\}$ and translate the relevant operations into a one-way computation. Subsequently the graphs for each of the individual operations can be combined to obtain the full ansatz in the one-way model. 

\begin{figure}[htb]
    \centering
    \includegraphics[width=0.8\columnwidth]{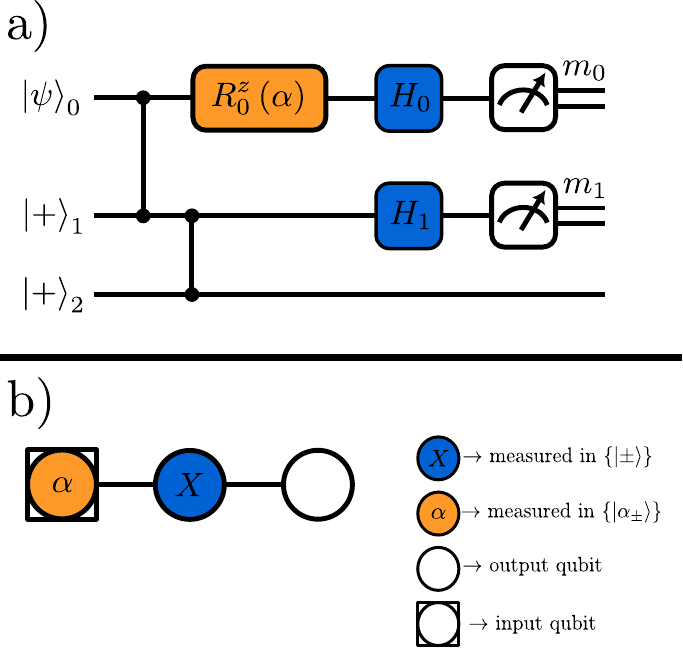}
    \caption{a) Single-qubit-teleportation-based scheme of the $z$-rotation $R^z(\alpha)$. b) The corresponding graph state for $\ket{\psi}_0=\ket{+}_0$ and measurement pattern for the one-way computation.}
    \label{fig:Rz-gate_sqtp_mbqc}
\end{figure}

Let us start with translating the single-qubit $z$-rotations into the one-way model. To this end, we will have to concatenate the single-qubit teleportation protocol of Fig.~\ref{fig:single_qubit_teleportation} two times, first between qubits $0$ and $1$ and second between qubits $1$ and $2$. The second time, we can omit the $z$-rotation, which means we have to measure qubit $1$ in the Pauli-$X$ eigenbasis. The whole teleportation scheme of a simple $z$-rotation and the corresponding one-way computation is shown in Fig.~\ref{fig:Rz-gate_sqtp_mbqc}. 
It is straightforward to show that the output state in qubit $2$ of Fig.~\ref{fig:Rz-gate_sqtp_mbqc}a) after both measurements is given by
\begin{equation}X^{m_1}Z^{m_0}R^z(\alpha)\ket{\psi}
\end{equation}
where the first two terms are Pauli corrections depending on measurement outcomes $m_i\in\{0,1\}$ of qubit $i$. As before, the combination of a $z$-rotation, a Hadamard gate and a measurement in the computational basis on the first qubit correspond to a measurement in the rotated $X$-basis, which is defined by the rotation angle $\alpha$ of the $z$-rotation. From this teleportation scheme type of circuit, we can directly extract the required graph state as well as the necessary measurement pattern. Figure~\ref{fig:Rz-gate_sqtp_mbqc}b) shows the representation as a graph for $\ket{\psi}_0=\ket{+}_0$. The two $CZ$ gates at the beginning again create a three-qubit linear graph state, where colored vertices indicate that the qubit is subsequently measured in the basis indicated by the label of the vertex. The vertex surrounded by a box indicates the input qubit, which is where the computation starts, and the vertex without filling indicates the output qubit containing the final state at the end of the computation.

\begin{figure}[h!]
    \centering
    \includegraphics[width=1.0\columnwidth]{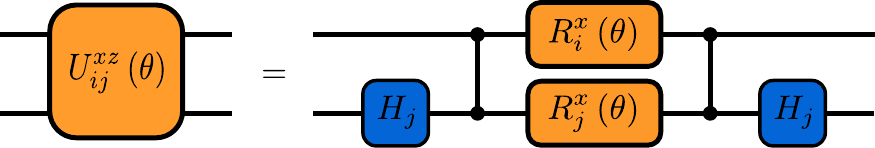}
    \caption{Efficient decomposition of $U^{xz}(\theta)$ in terms of $\{R^x,\,H,\,CZ\}$ gates.}
    \label{fig:xx-zz-gate_decomp}
\end{figure}

\begin{figure}[h!]
    \centering
    \includegraphics[width=1.0\columnwidth]{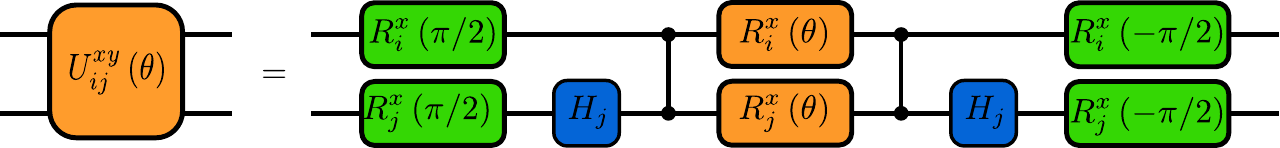}
    \caption{Decomposition of $U^{xy}(\theta)$ into the gate set $\{R^x,\,H,\,CZ\}$. The middle part part corresponds to the circuit for $U^{xz}(\theta)$ from Fig.~\ref{fig:xx-zz-gate_decomp}, which is padded with $R^x(\pm\pi/2)$ rotation gates as explained in the text.}
    \label{fig:xx-yy-gate_decomp}
\end{figure}

Next, we will decompose the entangling gate $U_{ij}^{xy}(\theta)$. Therefore, we employ an efficient decomposition of the unitary operation
\begin{equation}
    U^{xz}_{ij}(\theta) = e^{-i\theta/2(X_iX_j+Z_iZ_j)}
    \label{eq:xx-zz-gate}
\end{equation}
presented in Ref.~\cite{Smith2019} and shown in Fig.~\ref{fig:xx-zz-gate_decomp}.
Note that $U^{xz}_{ij}(\theta)$ can be transformed into $U^{xy}_{ij}(\theta)$ by applying local unitaries that change $Z_i Z_j$ to $Y_i Y_j$ while leaving $X_iX_j$ unchanged. This can be achieved with the operation $H\Phi H$ which transforms the Pauli matrices $X$, $Z$ as
\begin{align}
    (H \Phi H)^\dagger X (H \Phi H) &= -X,\\
    (H \Phi H)^\dagger Z (H \Phi H) &= Y
\end{align}
where $H$ is the usual Hadamard gate and $\Phi = e^{i\frac{\pi}{4}}R^z\left(\frac{\pi}{2}\right)$. Using that $H\Phi H = e^{i\frac{\pi}{4}}R^x\left(\frac{\pi}{2}\right)$, we can rewrite $U^{xy}_{ij}(\theta)$ in terms of single-qubit gates and $U^{xz}_{ij}(\theta)$ as shown in Fig.~\ref{fig:xx-yy-gate_decomp}.
If we arrange the operations $U^{xy}_{ij}(\theta)$ in a odd-even structure, as shown in Eq.~\eqref{eq:QuAlg} and Fig.\ \ref{fig:vqe_protocol_decomp}b), the last $x$-rotations $R^x(-\pi/2)$ of the even part cancel with the first $x$-rotations $R^x(\pi/2)$ of the odd part for all qubits, except the first and the last one. Hence, we do not need to reformulate operator $U^{xy}_{ij}(\theta)$ explicitly in the teleportation scheme, but instead we only reformulate the operator $U^{xz}_{ij}(\theta)$ and the general $x$-rotation $R^x(\alpha)$. From those two operations and the the general $z$-rotation in Fig.~\ref{fig:Rz-gate_sqtp_mbqc}, we can directly derive the one-way implementation of a whole layer of the ansatz. 

\begin{figure}[htp!]
    \centering
    \includegraphics[width=0.8\columnwidth]{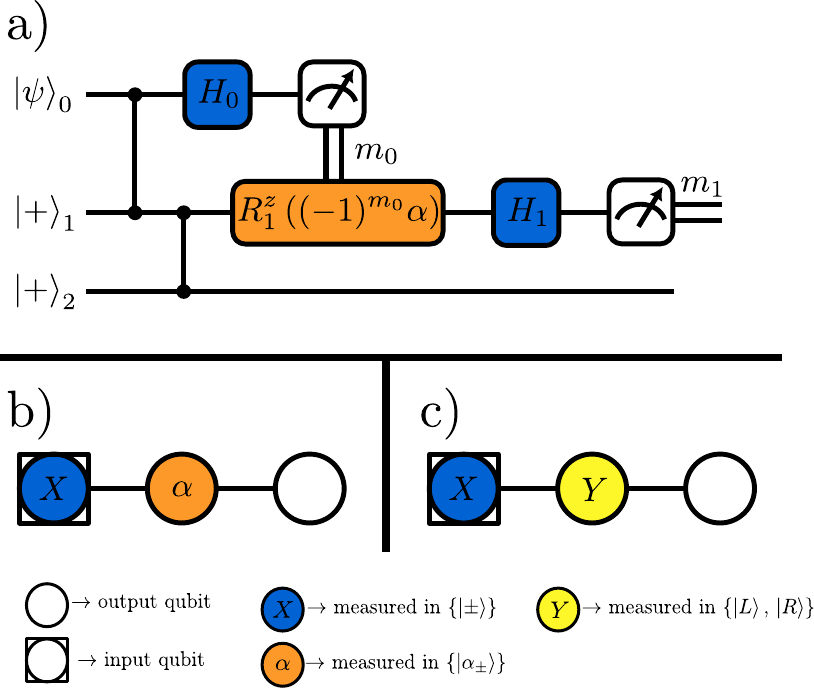}
    \caption{a) Single-qubit-teleportation-based scheme of the $x$-rotation gate $R^x(\alpha)=HR^z(\alpha)H$. b) The corresponding graph state for $\ket{\psi}_0=\ket{+}_0$ and measurement pattern for the one-way computation. c) Exemplary one-way computation for $\theta=-\pi/2$. In this case, the rotated basis $\{\alpha_\pm\}$ coincides with the Pauli-$Y$ eigenbasis $\{\ket{L},\,\ket{R}\}$ and the second measurement becomes non-adaptive.}
    \label{fig:Rx-gate_sqtp}
\end{figure}

The teleportation scheme of $R^x(\alpha)$ as well as the graph representation of the one-way model is shown in Fig.~\ref{fig:Rx-gate_sqtp}.
The first qubit is measured in the Pauli-$X$ eigenbasis, and the second qubit is measured in the rotated $X$-basis, $\{\ket{\theta_\pm}\}$. As illustrated in Fig.~\ref{fig:Rx-gate_sqtp}a) and \ref{fig:Rx-gate_sqtp}b), one has to change the sign of the measurement in the rotated $X$-basis based on the outcome of the Pauli-$X$ measurement, which corresponds to an adaptive measurement. The state of the third output qubit after both measurements are performed is given by
\begin{equation}
    X^{m_1}Z^{m_0}HR^z(\alpha)H\ket{\psi} = X^{m_1}Z^{m_0}R^x(\alpha)\ket{\psi},
    \label{eq:Rx_mbqc_output}
\end{equation}
where two leftmost Pauli operations in the equation above depend on the measurement outcomes $m_0,\,m_1\in\{0,1\}$ of the first two qubits. Note that, Fig.~\ref{fig:Rx-gate_sqtp}b) looks very similar to Fig.~\ref{fig:Rz-gate_sqtp_mbqc}b) with only the measurement basis of qubits $0$ and $1$ being interchanged, which in turn makes the rotated $X$-basis measurement adaptive. The adaptive choice is indicated by the possible minus sign of the rotated basis angle depending on the measurement outcome $m_0$ in Fig.~\ref{fig:Rx-gate_sqtp}a). The required graph state with the necessary measurement pattern is shown in Fig.~\ref{fig:Rx-gate_sqtp}b). Note that for $\alpha=-\pi/2$ the rotated basis $\{\ket{\alpha_\pm}\}$ coincides with the Pauli $Y$ eigenbasis $\{\ket{L},\ket{R}\}$ and that $R^x(\pi/2)=XR^x(-\pi/2)$. Hence, for the implementation of $R^x(\pm\pi/2)$ the rotated basis measurement becomes a non-adaptive Pauli $Y$ measurement with an additional Pauli $X$ correction for $R^x(+\pi/2)$ (cf.\ Fig.~\ref{fig:Rx-gate_sqtp}c)).

\begin{figure*}
    \centering
    \includegraphics[width=1.0\textwidth]{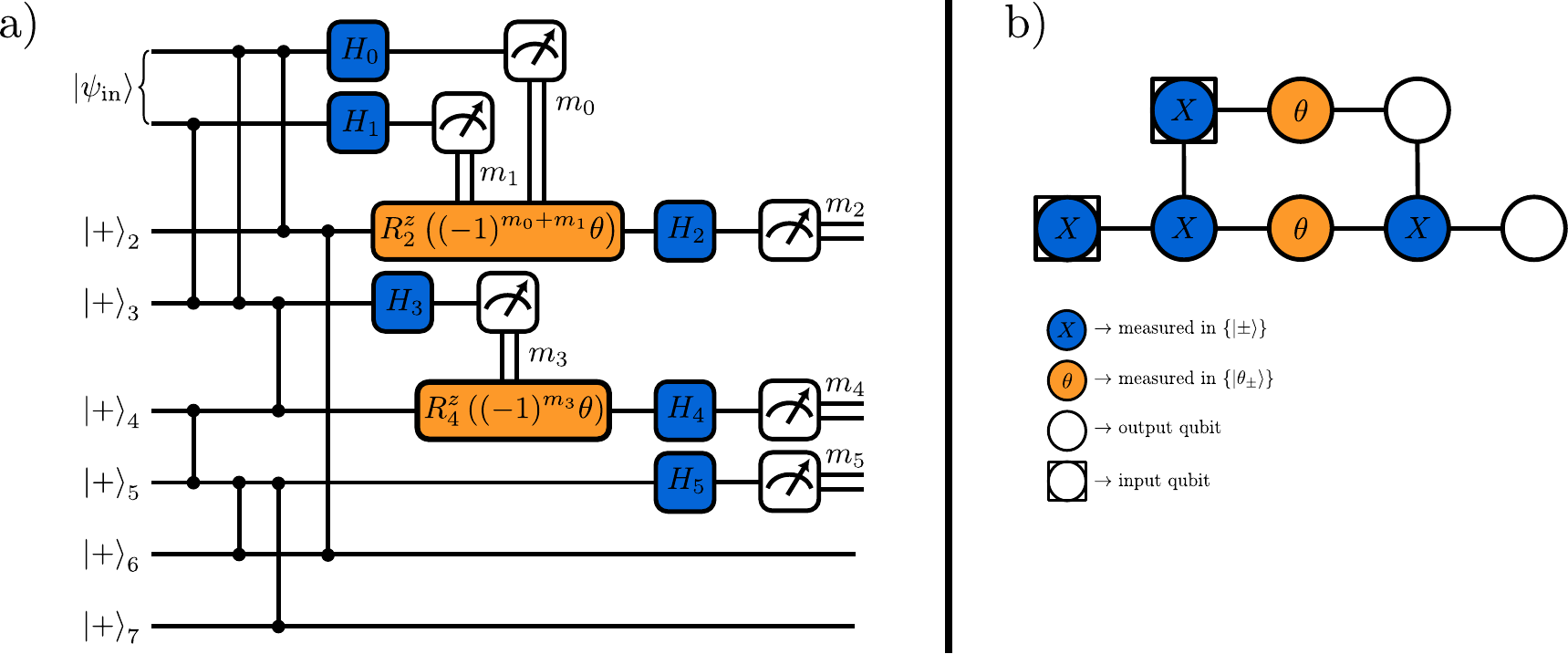}
    \caption{a) Single-qubit-teleportation-based scheme of $U^{xz}(\theta)$, shown in Eq.~\eqref{eq:xx-zz-gate}. b) The corresponding graph state for $\ket{\psi_{\mathrm{in}}}_{01}=\ket{+}_0\ket{+}_1$ and measurement pattern for the one-way computation.}
    \label{fig:xx-zz-gate_sqtp}
\end{figure*}

Next, the teleportation scheme of $U^{xz}_{ij}(\theta)$ is shown in Fig.~\ref{fig:xx-zz-gate_sqtp}, together with the corresponding one-way computation.
The final state of the output qubits $6$, $7$ in Fig.~\ref{fig:xx-zz-gate_sqtp}a) is given by
\begin{equation}
\ket{\psi_{\mathrm{out}}}_{67} = U_{\mathrm{PC}}U^{xz}_{67}(\theta)\ket{\psi_{\mathrm{in}}}_{67},
\end{equation}
with the Pauli correction 
\begin{equation}
    U_{\mathrm{PC}} = \left(X^{m_2}Z^{m_0+m_4}\right)_6\left(X^{m_5+m_2+m_3}Z^{m_4+m_1}\right)_7.
\end{equation} 
The one-way computation of the whole ansatz layer can be obtained by concatenating the individual one-way computations of $U^{xz}_{ij}(\theta)$, $R^x(\pm\pi/2)$ and $R^z(\theta)$ according to Eq.~\eqref{eq:QuAlg} and Fig.~\ref{fig:xx-yy-gate_decomp}. 

\begin{figure*}
    \centering
    \includegraphics[width=0.8\textwidth]{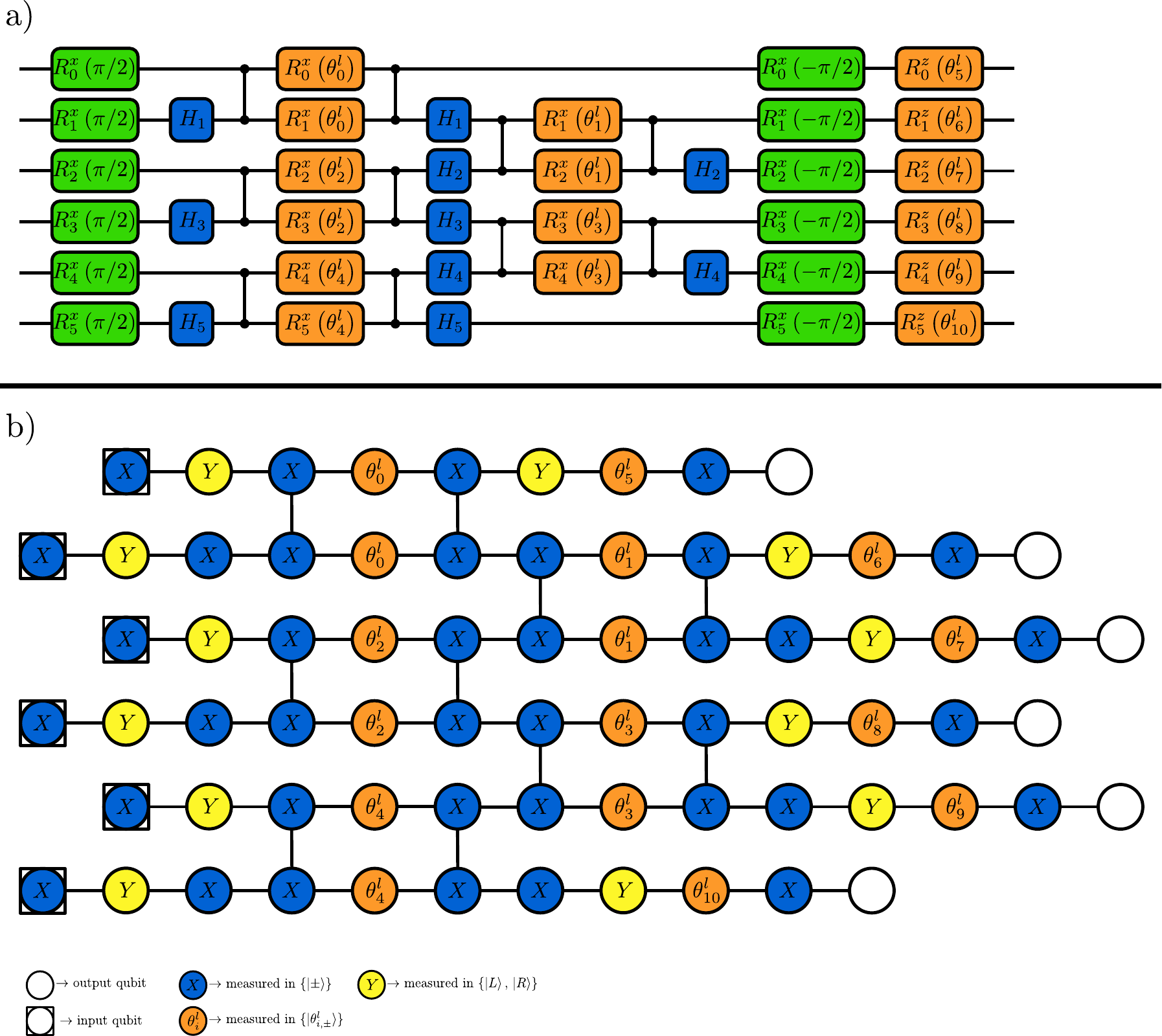}
    \caption{a) Decomposition of the ansatz layer, shown in Eq.~\eqref{eq:QuAlg}, for 6 qubits in the set $\{H,\, R^z,\, R^x,\, CZ\}$. b) Implementation of the ansatz layer for 6 qubits in the one-way model.}
    \label{fig:ansatz_decomp}
\end{figure*}

Finally, Fig.~\ref{fig:ansatz_decomp} shows our algorithm layer for six input qubits in the gate set $\{H,\,R^z,\,R^x,\,CZ\}$ (cf.\ Fig.~\ref{fig:ansatz_decomp}a)) and the corresponding one-way implementation (cf.\ Fig.~\ref{fig:ansatz_decomp}b)).
In Fig.~\ref{fig:ansatz_decomp}b), the adaptive basis choice and the resulting Pauli corrections of the whole algorithm layer can be obtained by concatenating the output states of the individual sub-operations in the one-way scheme accordingly and than permute all Pauli corrections to the outer left. Note that in general all non-adaptive measurements in a one-way quantum computation can be performed in parallel and in the first step of the computation. All adaptive measurements have to be performed in a temporal order prescribed by the measurement dependencies in the adaptive basis choice.
The number of qubits $Q_{\mathrm{owqc}}$ in the one-way computation of the whole ansatz layer scales with the number of input qubits $NF$ (qubits on which the ansatz layer is applied) as 
\begin{equation}
    Q_{\mathrm{owqc}} = 13\times NF-6.
\end{equation}
Thus, a large number of qubits is required to implement our ansatz layer in a one-way computation, which is currently beyond any physical realization of graph states.
In principle, one could reduce the number of qubits by using the fact that the non-adaptive Pauli measurements in a one-way quantum computation correspond to a Clifford operations~\cite{Ferguson:2020qyf, Marqversen2023}. Hence, they can all be performed in parallel in the first step of the computation~\cite{Ferguson:2020qyf, Marqversen2023}. Furthermore, instead of performing those Pauli measurements on the graph state, they can also be efficiently simulated classically beforehand thanks to the Gottesman-Knill Theorem~\cite{Ferguson:2020qyf}. More specifically, since all Pauli measurements on a graph state yield a state which is local-unitary-equivalent (LU-equivalent) to another graph state, the classical simulation of Pauli measurements on graph states can be reformulated in a set of graph modification rules. These transform the initial graph state to the graph state LU-equivalent to the state after the Pauli measurements~\cite{Hein2004, Kok2010}. Since all the previously measured qubits are disentangled from the resulting graph state, this procedure reduces the size of the required graph state in the experiment. On the resulting state, only the adaptive measurements (non-Clifford part of the computation) have to be executed. However, as pointed out in Ref.~\cite{Marqversen2023}, the resulting graph states get additional (possibly long-ranged) $CZ$ connections between the remaining qubits, which destroy the geometric structure within the graph representation. While such unstructured graph states require a smaller number of vertices and correspondingly a smaller number of qubits, they become increasingly complicated to realize experimentally with growing system size due the unstructured $CZ$ connections.

\section{Results\label{sec:results}}

To benchmark our ansatz, we consider the Schwinger model with three fermion flavors with flavor-dependent chemical potentials. We simulate our VQE protocol classically assuming a perfect quantum computer without any hardware or shot noise, and compare the results to the ones obtained from exact diagonalization. As a classical optimizer for the VQE, we choose the L-BFGS algorithm~\cite{Nocedal2006}. For each value of the chemical potential, ten different runs of the simulation with randomly chosen initial parameters are carried out. Subsequently we post-process our data, and mark simulations with a final energy that is 30\% higher than the lowest value obtained within the ten runs as outliers.  To assess the performance of our ansatz, we study various parameter regimes, in particular, we investigate (i) vanishing bare fermion mass (ii) non-vanishing bare fermion mass, and (iii) a sign-problem afflicted regime for conventional MC simulations.
In addition, to demonstrate the feasibility of our approach on quantum hardware, we also carry out inference runs for six qubits (corresponding to $N=2$ for $F=3$) on an actual IBM quantum hardware. To this end, we take the optimal variational parameters obtained from the classical simulations for two ansatz layers and execute a pre-transpiled version of our ansatz with those parameters, measuring the relevant observables directly on the quantum device.

Since our primary focus lies on investigating the performance of our ansatz for the lattice system and do not intend to take the continuum limit, we will not take the additive mass renormalization into account in our simulations. 

\subsection{Vanishing bare fermion mass}
\begin{figure*}
  \centering
  \includegraphics[width=1.0\textwidth]{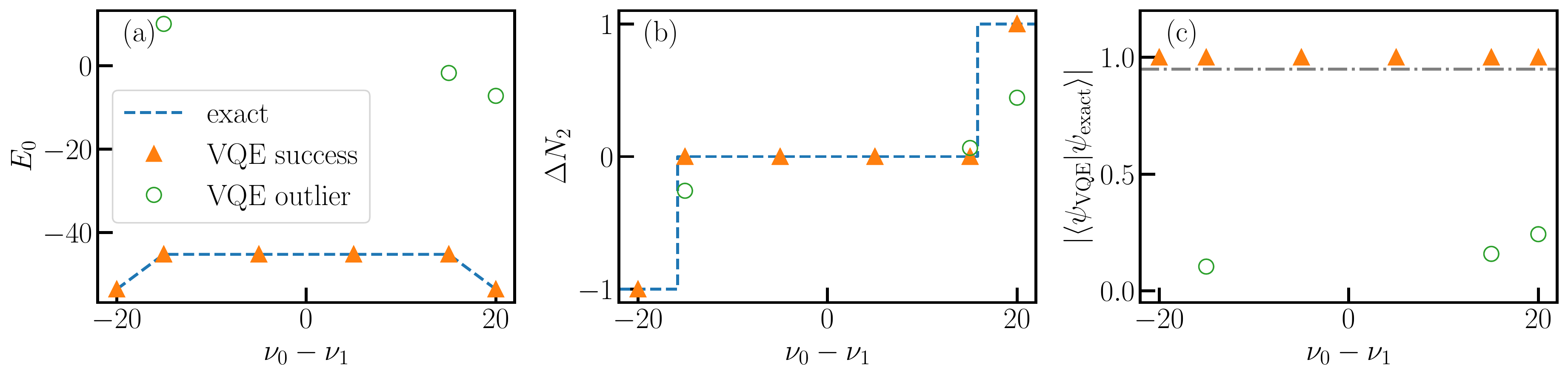}
  \includegraphics[width=1.0\textwidth]{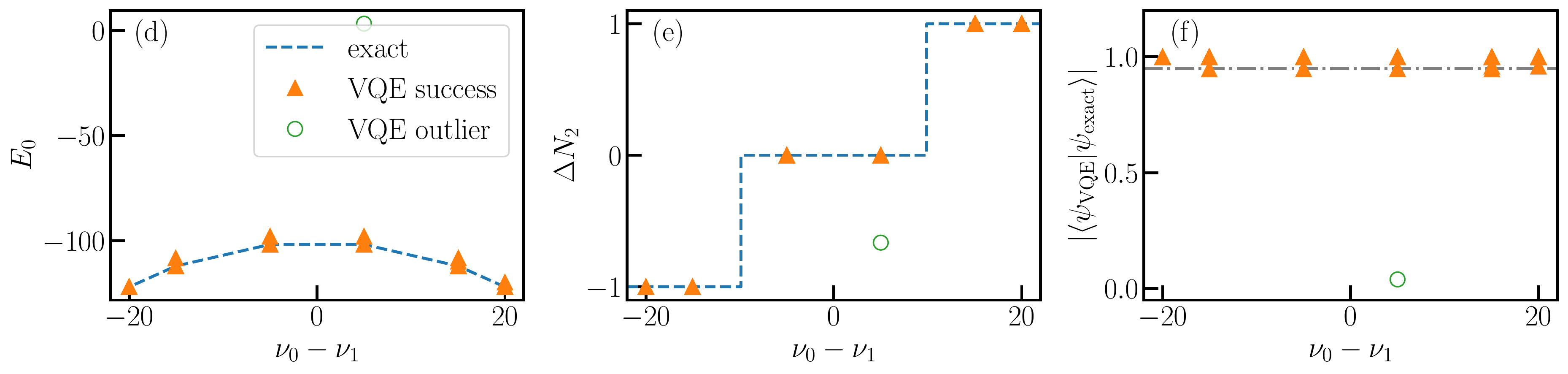}
  \includegraphics[width=1.0\textwidth]{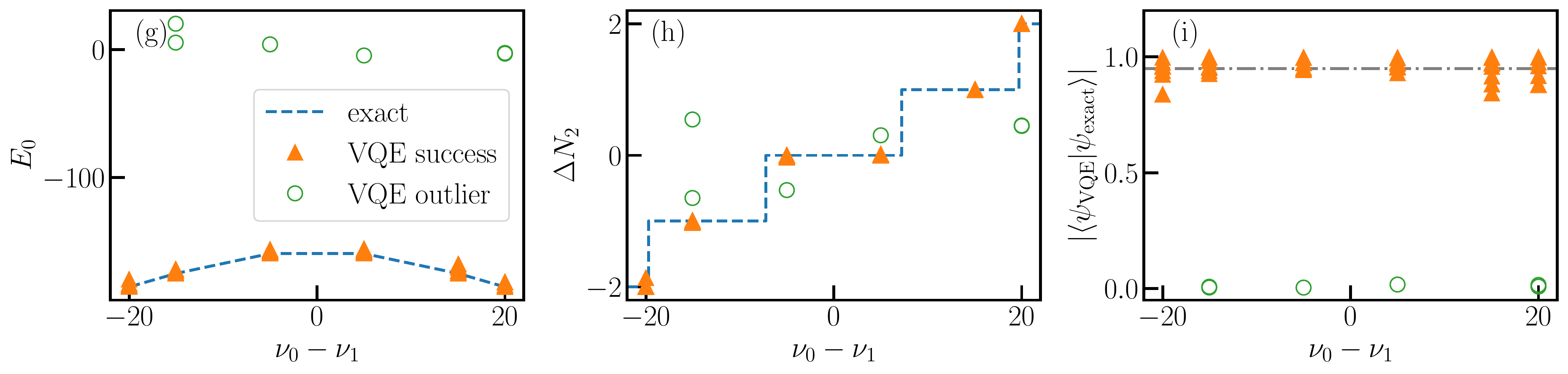}
  \caption{Classically simulated VQE results for vanishing bare fermion mass. The first column shows the ground-state energy $E_0$, the second column shows the particle number difference $\Delta N_2$, and the third column shows the overlap $\lvert{\braket{\psi_\mathrm{VQE}|\psi_\mathrm{exact}}}\lvert$ of the VQE results, each as a function of the chemical potential difference $\nu_0-\nu_1$. The first row shows the results for $N=2$, the second row for $N=4$ and the third row for $N=6$, each for $x=16$, $\mu_f=0$, $\nu_2=-\nu_0$, $\nu_1=0$, and five algorithm layers. Successful runs are represented by filled orange triangles. Every run, in which the obtained energy is 30\% higher than the lowest obtained value, is marked as an outlier and is represented by an open green circle. The solution obtained via exact diagonalization is shown as a dashed blue line in the first two columns. The dash-dotted grey line in the third column marks the 95\% overlap threshold.}
  \label{fig:N2m0}
\end{figure*}

To begin with, we focus on the case of vanishing bare fermion mass, $\mu_f=0$, and consider $\nu_2 = -\nu_0$ with $\nu_1=0$. Thus, we are in a regime in which the model has the reflection symmetry discussed in Sec.~\ref{sec:model}, allowing us to constrain the parameters in the ansatz according to Eqs.~\eqref{eq:QuAlgSymmRest_oe} and \eqref{eq:QuAlgSymmRest_z}. Figure~\ref{fig:N2m0} shows our results for the ground state energy, the particle number and the overlap obtained by simulating the VQE classically for system sizes $N=2$, $4$ and $6$, which correspond to $6$, $12$ and $18$ qubits respectively.
For the entire range of system sizes and chemical potentials we study, we generally observe good agreement between the VQE results and the exact solution computed with exact diagonalization. Focusing on the overlap with the exact solution in Figs.~\ref{fig:N2m0}(c), \ref{fig:N2m0}(f) and \ref{fig:N2m0}(i), we see that in most cases we achieve overlaps with the exact ground state of more than $95\%$, and the values only decrease marginally with increasing system size, although we use a constant number of $5$ layers in the ansatz for all system sizes. The cases in which the overlap is significantly lower than $95\%$ correspond to outliers according to our criterion that the energy is at least $30\%$ higher than the lowest one obtained within the ten runs, as the panels for the energy in the first column of Fig.~\ref{fig:N2m0} reveal. While the exact ground state energy as a function of the chemical potential already indicates the onset of first-order quantum phase transitions in form of cusps, these are harder to detect from the VQE results for the energy, as one would need an extremely fine resolution in the chemical potential. In contrast, the first-order quantum phase transitions manifest themselves clearly in the particle number in form of characteristic jumps when going from one phase to another, as the panels in the middle column of Fig.~\ref{fig:N2m0} show. The different phases are all well captured by the VQE results and can be reliably identified with a modest number of data points.

Looking at the outliers in Fig.~\ref{fig:N2m0}, we see that these can be easily identified via the energy and the particle numbers, observables which can be efficiently measured on actual quantum hardware. They consistently show non-integer particle numbers and high energy values, giving a strong indication that they are unphysical. This is confirmed by the almost vanishing overlaps with the exact ground states. Moreover, within our ten experiments only a small fraction of simulations produced outliers, showing that that our setup is very likely to produce a good approximation for the ground state of the model. Consequently, in the regime of vanishing bare fermion mass our VQE protocol is able to reliably capture the phase structure of the model, and occasional outliers can be determined easily from the observables considered.

\subsection{Non-vanishing bare fermion mass}
\begin{figure*}
  \centering
  \includegraphics[width=1.0\textwidth]{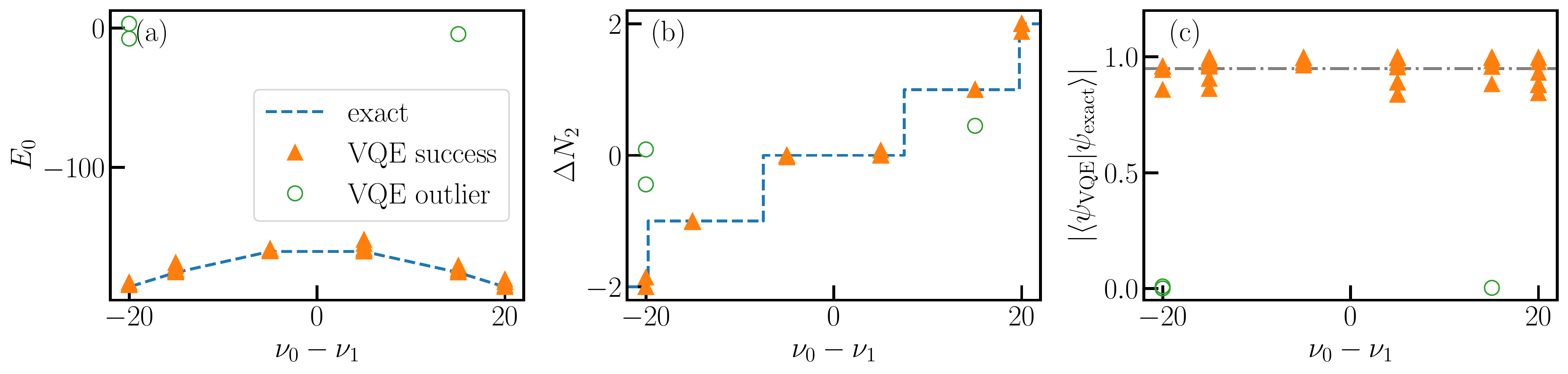}
  \caption{Classically simulated VQE results for non-vanishing bare fermion mass. Panel (a) shows the ground-state energy $E_0$, panel (b) the particle number difference $\Delta N_2$, and panel (c) the overlap $\lvert{\braket{\psi_\mathrm{VQE}|\psi_\mathrm{exact}}}\lvert$ of the VQE results, each as a function of the chemical potential difference $\nu_0-\nu_1$ for $N=6$, $x=16$, $\mu_f=0.1$, $\nu_2=-\nu_0$, $\nu_1=0$, and five algorithm layers. Successful runs are represented by filled orange triangles. Every run, in which the obtained energy is 30\% higher than the lowest obtained value, is marked as an outlier and is represented by an open green circle. The solution obtained via exact diagonalization is shown as a dashed blue line in the first two panels. The dash-dotted grey line in the third panel marks the 95\% overlap threshold.}
  \label{fig:N6m01}
\end{figure*}

As a next step, we consider a nonzero bare fermion mass of $\mu_f=0.1$ while still keeping $\nu_1=0$ and $\nu_2=-\nu_0$. Moreover, we focus on the largest system size we studied in the previous section, $N=6$ corresponding to $18$ qubits. Figure~\ref{fig:N6m01} shows our VQE results for the ground-state energy and the particle number in comparison to the results from exact diagonalization as well as the overlap with the exact solution.
Focusing on the exact results energy and the particle number in Figs.~\ref{fig:N6m01}(a) and \ref{fig:N6m01}(b), we observe qualitatively the same behavior as for the case of vanishing fermion mass, the particle number shows again abrupt discontinuities indicating the first-order phase transitions which are accompanied by cusps in the energy. Our VQE results for the energy and the particle number are in general in good agreement with the exact solution, which is also corroborated by looking at the overlap between the VQE solution and the exact wave function in Fig.~\ref{fig:N6m01}(c). Again, we are able to obtain overlaps that are around $95\%$ for most of the simulations we carry out. 

Similar to the previous case of vanishing bare fermion mass, for $\mu_f=0.1$ we also see some data points that are identified as outliers according to our criterion. A direct comparison between Figs.~\ref{fig:N2m0} and \ref{fig:N6m01} shows that for $\mu_f=0.1$ we observe an even smaller fraction of simulations that produced outliers than for $\mu_f=0$, and our VQE converges with high probability. These outliers can again reliably be identified by looking at the physical observables, and manifest themselves in high values for the energy and noninteger particle numbers. This is likely caused by the classical optimization routine getting stuck in a local minimum, as in these cases the final VQE wave function has almost vanishing overlap with the exact wave function, despite the ansatz being capable of approximating it to a good precision, as the other successfull runs demonstrate.

\subsection{Sign-problem afflicted regime}\label{sec:results_sign_prob}
\begin{figure*}
  \centering
  \includegraphics[width=1.0\textwidth]{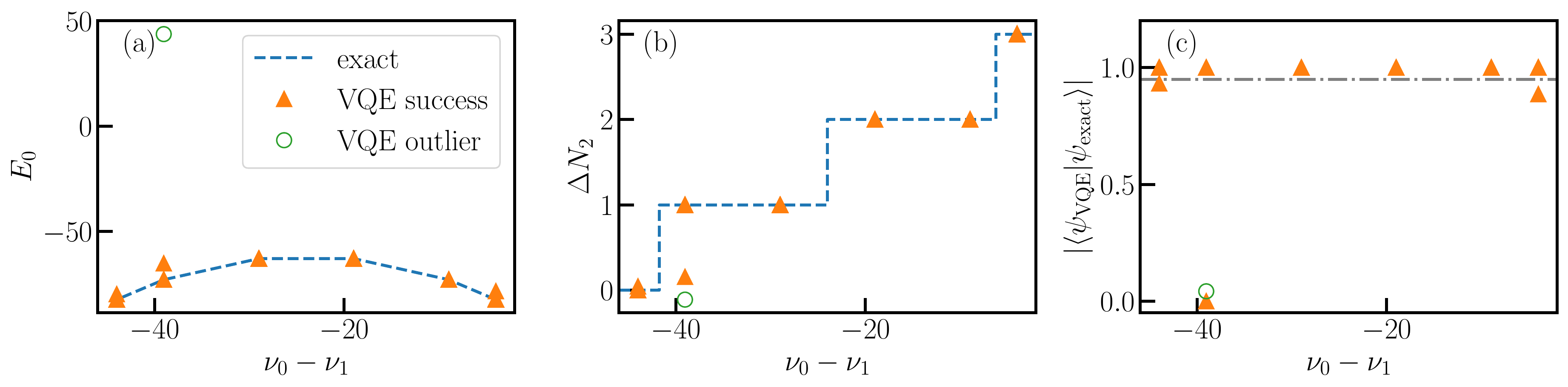}
  \caption{Classically simulated VQE results for sign-problem afflicted regime. Panel (a) shows the ground-state energy $E_0$, panel (b) shows the particle number difference $\Delta N_2$, and panel (c) shows the overlap $\lvert{\braket{\psi_\mathrm{VQE}|\psi_\mathrm{exact}}}\lvert$ of the VQE results, each as a function of the chemical potential difference $\nu_0-\nu_1$ for $N=4$, $x=16$, $\mu_f=0.1$, $\nu_2=-\nu_0$, $\nu_1=24$, and five layers in the ansatz. Successful runs are represented by filled orange triangles. Every run, in which the obtained energy is 30\% higher than the lowest obtained value, is marked as an outlier and is represented by an open green circle. The solution obtained via exact diagonalization is shown as a dashed blue line in the first two panels. The dash-dotted grey line in the third panel marks the 95\% overlap threshold.}
  \label{fig:N4m0_kg3}
\end{figure*}

Next, we investigate a regime which is inaccessible with MC methods, due to the sign problem. To this end, we consider again $\mu_f=0$ and $\nu_2=-\nu_0$, but now we additionally set $\nu_1=24$. This results in $\sum_f\nu_f\neq 0$, thus triggering a sign problem for the conventional MC approach. In addition, for this case the reflection symmetry of the model is no longer present, and we cannot constrain the variational parameters in the ansatz anymore and treat all the parameters in the ansatz as independent. Once again, we investigate this regime for a system with 12 qubits, corresponding to four lattice sites.
Figure~\ref{fig:N4m0_kg3} shows the results obtained for the simulations in this parameter regime.
Our VQE ansatz still produces results, which are in good agreement with the exact solution, even though conventional MC methods would fail in this situation. Because we can no longer constrain the number of parameters in our ansatz, we get now $23$ parameters per layer instead of $12$ as before for $N=4$. Nevertheless, the optimization is still capable of finding a good approximation of the exact ground state, and we do not see a significant increase in the number of outliers, comparing Figs.~\ref{fig:N2m0}(f) and \ref{fig:N4m0_kg3}(c). We do observe a single data point which has vanishing overlap with the exact ground state, but is not marked as a outlier since its energy is close to the exact one and the particle number is approximately an integer value. This data point lies close to a phase transition point, at which the ground-state energy level is degenerate, which means that the energy levels of the ground states from each neighbouring phase are still close. It is likely that the VQE converged to the wrong ground state, which would explain why the energy is close to the exact solution but the particle number differs and the overlap vanishes. In general, our ansatz still achieves high overlaps with the exact solutions for a large fraction of the simulation runs and shows overall a good performance, even in this sign-problem afflicted regime.

\subsection{Inference run on IBM quantum hardware}
\label{sec:results_ibm}
\begin{figure*}
  \centering
  \includegraphics[width=1.0\textwidth]{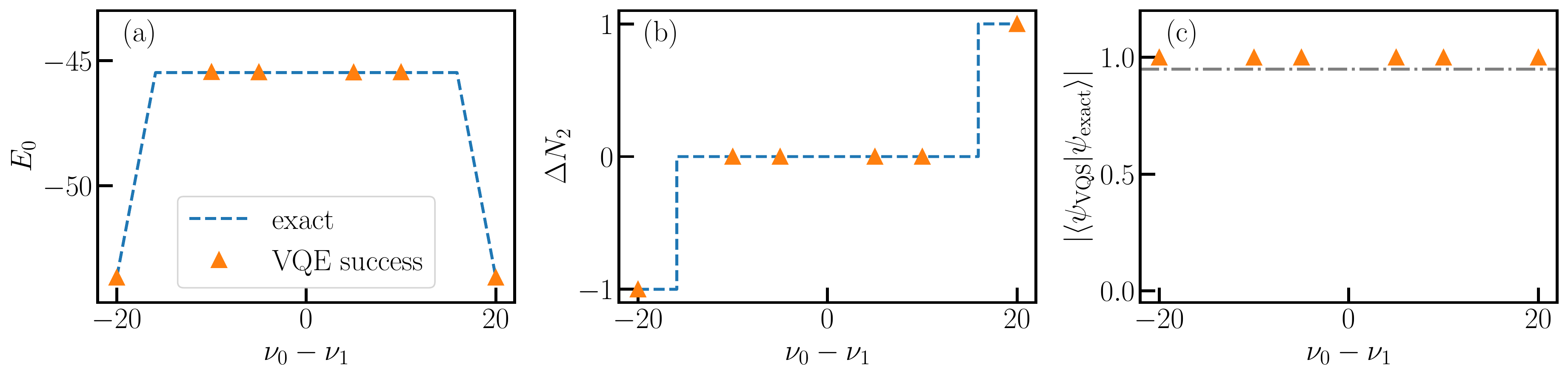}
  \caption{Classically simulated VQE results for inference run on ibm\_cairo. The panel (a) shows the ground-state energy $E_0$, the panel (b) shows the particle number difference $\Delta N_2$ and the panel (c) shows the overlap $\lvert{\braket{\psi_\mathrm{VQE}|\psi_\mathrm{exact}}}\lvert$ of the VQE results, each as a function of the chemical potential difference $\nu_0-\nu_1$ for $N=2$, $x=16$, $\mu_f=0.1$, $\nu_2=-\nu_0$, $\nu_1=0$, and two algorithm layers. Successful runs are represented by filled orange triangles. The solution obtained via exact diagonalization is shown as a dashed blue line in the first two panels. The dash-dotted grey line in the third panel marks the 95\% overlap threshold.}
  \label{fig:N2m01_infer_vqe}
\end{figure*}

To demonstrate that our ansatz can be implemented on current NISQ devices, we perform inference runs on real quantum hardware and directly measure the relevant observables on the quantum computer. Specifically, we prepare our ansatz with two layers using the optimal parameters obtained from the classical simulations in the regime of non-vanishing bare fermion mass on  ibm\_cairo. We focus on the case of six qubits, corresponding to $N=2$ lattice sites for three fermion flavors, for which we obtained essentially perfect agreement with the exact solution with overlaps exceeding $99\%$ with the exact ground state in the classical simulation, as shown in Fig.~\ref{fig:N2m01_infer_vqe}. 

\begin{figure*}
    \centering
    \includegraphics[width=1.0\textwidth]{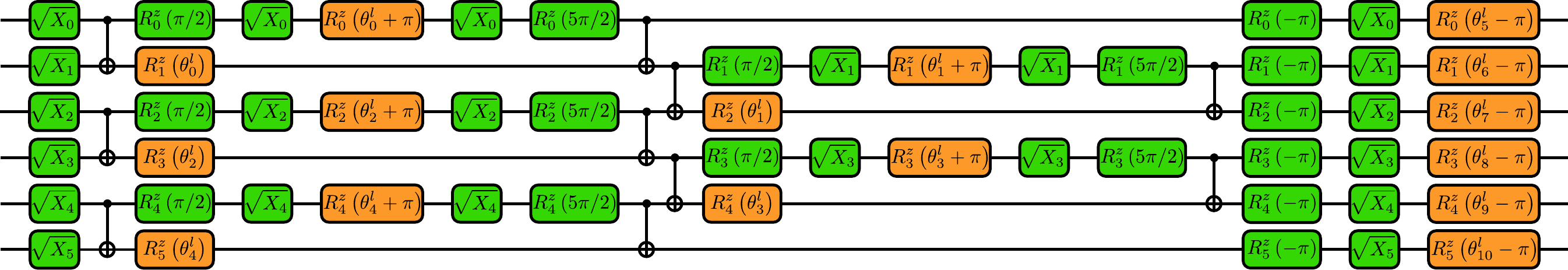}
    \caption{Illustration of the decomposition of a full layer of the ansatz (see Eq.~\eqref{eq:QuAlg}) into the native gate set $\{CX,\, R^z,\, \sqrt{X},\, X\}$ of ibm\_cairo  for six qubits, corresponding to $N=2$, $F=3$.}
    \label{fig:ansatz_decomp_ibm}
\end{figure*}

For the hardware run, we decompose our ansatz circuit into the natively available set of gates on ibm\_cairo, $\{CX,\,R^z,\,\sqrt{X},\, X\}$. To this end, we employ the efficient decomposition of our ansatz layers (see Fig.~\ref{fig:ansatz_decomp}) and further decompose those gates into the native basis gate set, resulting in the structure shown in Fig.~\ref{fig:ansatz_decomp_ibm}. 
In addition, we eliminate all two-qubit gates in the ansatz which result in operations close to the identity for a given parameter set. Therefore, we checked if the parameter value for the two-qubit gate agrees within $0.18\,\mathrm{rad}$ accuracy with $m\times2\pi\,\mathrm{rad}$, $m\in \mathbb{Z}$. If that is the case, the corresponding two-qubit gate is considered as a close-to-identity operation and removed from the circuit. 

\begin{figure*}[htb]
  \centering
  \includegraphics[width=1.0\textwidth]{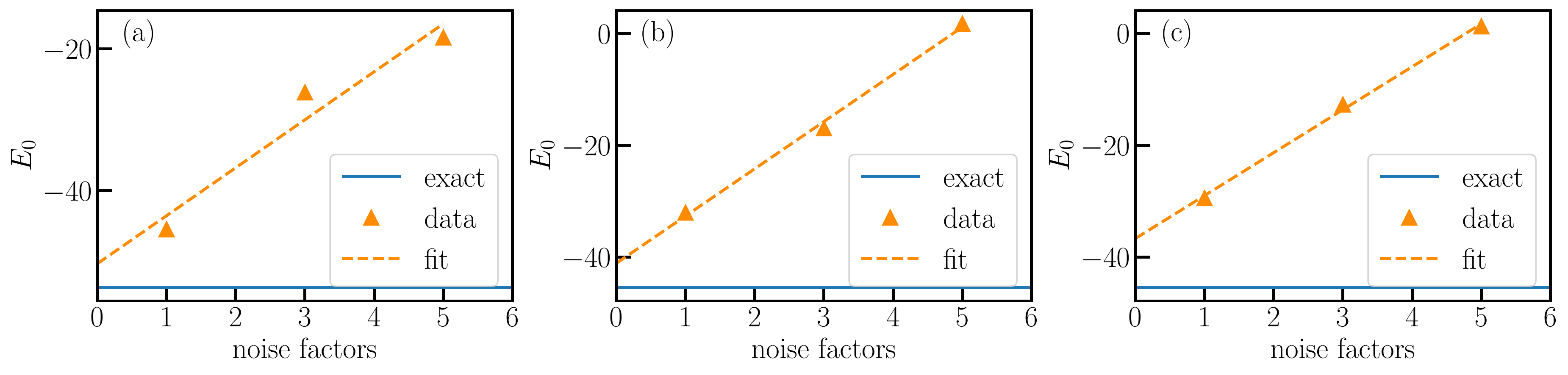}
  \includegraphics[width=1.0\textwidth]{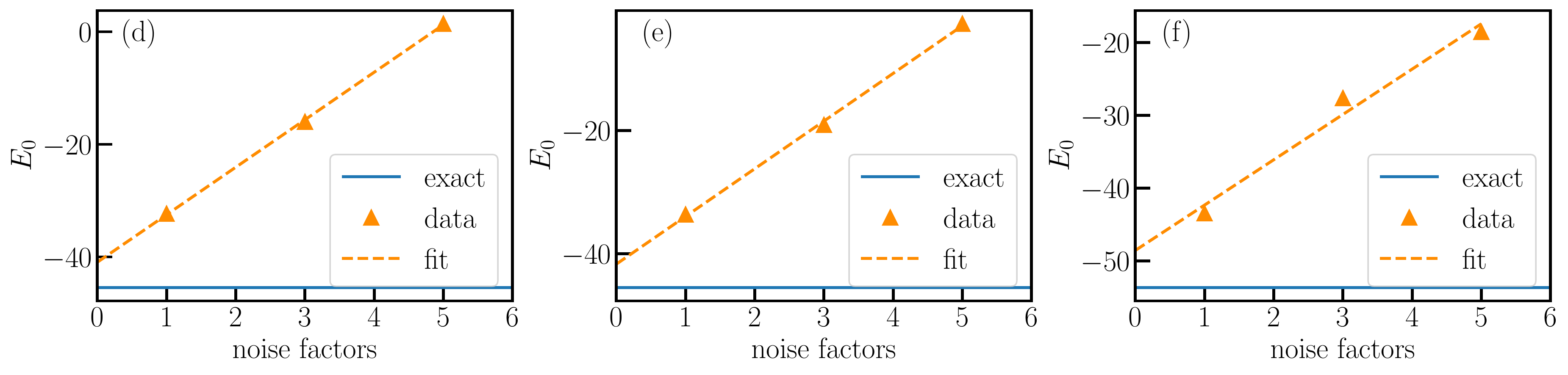}
  \caption{Zero-noise extrapolation of the ground-state energy from the inference run on ibm\_cairo. The filled orange triangles correspond to the hardware data for different noise amplification factors. The dashed orange lines show the linear extrapolation. The solid blue lines indicate the exact solutions obtained via exact diagonalization. Panel (a) to (f) correspond to the different chemical potentials $\nu_0-\nu_1=-20$, $-10$, $-5$, $5$, $10$, $20$, each with 15360 measurement shots, $N=2$, $x=16$, $\mu_f=0.1$, $\nu_2=-\nu_0$, $\nu_1=0$, and two algorithm layers. The noise factors indicate how often the original circuit, represented by a unitary $U$, was folded, i.e.\ noise factor 1 corresponds to the original circuit, noise factor 3 to $U(U^\dagger U)$ and noise factor 5 to $U(U^\dagger U)(U^\dagger U)$.}
  \label{fig:N2m01_hardware_zne}
\end{figure*}

In order to compensate for some of the hardware noise in the remaining circuit, we use zero-noise extrapolation (ZNE)~\cite{Giurgica-Tiron2020} available in Qiskit Runtime~\cite{qiskit_runtime_error_mitigation_doc}. The ZNE in Qiskit Runtime employs digital circuit folding, i.e.\ inserting pairs of unitaries $U^\dagger U$ in the circuit which would result in an identity operation on an ideal quantum computer. For a real device with noise, this effectively allows for running the same circuit at different noise levels. Subsequently, the results can be extrapolated to zero noise. Figure~\ref{fig:N2m01_hardware_zne} shows the data for the energy obtained on the hardware for various noise levels. In general, we observe the data points from the hardware do follow a linear behavior as a function of the noise level for all the values of the chemical potential we study. Performing a linear extrapolation significantly improves energy values, but does not fully compensate for the effects of noise, as the final extrapolated value for the energy is still above the exact one (cf.\ the solid blue lines and the dashed orange lines at the origin of the panels in Fig.~\ref{fig:N2m01_hardware_zne}).

\begin{figure*}
  \centering
  \includegraphics[width=1.0\textwidth]{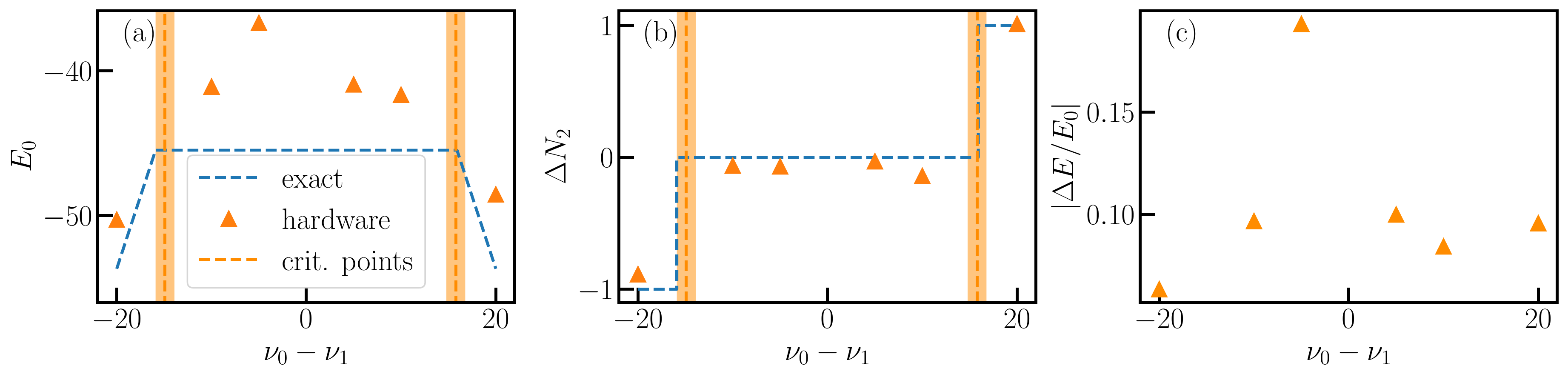}
  \caption{Quantum hardware results for VQE inference run on ibm\_cairo. The panel (a) shows the ground-state energy $E_0$, the panel (b) shows the particle number difference $\Delta N_2$ and the panel (c) shows the relative energy error $\lvert\Delta E/E_0\lvert$ of the inference run on ibm\_cairo with $15360$ measurement shots as a function of the chemical potential difference $\nu_0-\nu_1$ for $N=2$, $x=16$, $\mu_f=0.1$, $\nu_2=-\nu_0$, $\nu_1=0$, and two algorithm layers. Hardware runs are represented by filled orange triangles. The solution obtained via exact diagonalization is shown as a dashed blue line in the first two panels. The dashed orange lines together with the shaded orange area show the critical points and their standard deviation obtained from the hardware results via Eq.~\eqref{eq:nujump_3F_dN}.}
  \label{fig:N2m01_hardware}
\end{figure*}

The results obtained from the quantum hardware for the energy as well as the particle number after performing the ZNE as a function of the chemical potential are shown Fig.~\ref{fig:N2m01_hardware}. 
Focusing on the energy in Fig.~\ref{fig:N2m01_hardware}(a), we clearly see a systematic offset $\Delta E$ between the hardware results compared to the exact solution, as already apparent in the ZNE data in Fig.~\ref{fig:N2m01_hardware_zne}. Compared to exact value of the energy, this offset is rather modest in almost all cases, and the relative error $|\Delta E / E_0|$ is within $10\%$, as Fig.~\ref{fig:N2m01_hardware}(c) shows. Only in one of our runs we have an outlier with slightly higher relative error of around $17\%$. These deviations are clearly the result of the hardware noise, since we obtained perfect agreement between the exact solution and the VQE results and the ideal simulations (cf.\ Fig.~\ref{fig:N2m01_infer_vqe}). 

Looking at the particle number in Fig.~\ref{fig:N2m01_hardware}(b), we see that the hardware data agrees well with the exact result. In particular, the particle number clearly reveals the first-order phase transitions with the characteristic jumps being evident in the data from the quantum hardware. Note, that ZNE was also applied to the measurement of the particle numbers, in the same way as for the energy measurement. Moreover, the remaining error in the measured particle numbers can be compensated completely. Since the Hamiltonian conserves the particle number, only integer values are intrinsically allowed. Since all our data points within one phase are close to the same integer, we can reliably determine the exact values by rounding them to the nearest integer.

From our hardware data we can also determine the location of the first-order phase transitions via Eq.~\eqref{eq:nujump_3F_dN}. As shown in Sec.~\ref{sec:phasestructure}, a single data point per phase is in principle enough to determine the exact location of the phase boundaries. In order to average over the fluctuations between different runs in the phase characterized with $\Delta N_2 = 0$, we consider all possible pairs of data points for neighboring phases and determine the transition point according to Eq.~\eqref{eq:nujump_3F_dN} for all of them. Subsequently, we average the results and compute the standard deviation, the results are indicated in Figs.~\ref{fig:N2m01_hardware}(a) and \ref{fig:N2m01_hardware}(b) as vertical dashed orange lines. Note, that for every calculation of the phase transition points we fully compensated the error in the particle numbers by rounding to the nearest integer as we explained above, leaving only the energy error to contribute in the calculations. The numerical values obtained from our hardware run are $\nu_0-\nu_1\lvert^i_{\mathrm{jump},0}=-14.91\pm0.99$ and $\nu_0-\nu_1\lvert^i_{\mathrm{jump},1}=15.78\pm0.99$. They are close to the exact transition points $\nu_0-\nu_1\lvert^e_{\mathrm{jump},0,1}=\pm 15.91$. We observe that the value extracted for the second transition point $\nu_0-\nu_1\lvert^i_{\mathrm{jump},1}$ shows a smaller deviation from the exact value than the one for the first transition point $\nu_0-\nu_1\lvert^i_{\mathrm{jump},0}$. This likely due to similar energy offsets in the hardware results for points in the phases with $\Delta N_2 = 0$ and $\Delta N_2 = 1$. As Eq.~\eqref{eq:nujump_3F_dN} reveals, adding the same constant offset to both energy values in the formula does not affect the result. All in all, our data from the quantum hardware allows us to determine the critical points of the model with good accuracy, despite the presence of noise in the device.

\section{Discussion and Conclusions}
\label{sec:conclusion}
We presented a VQE ansatz for solving the lattice multi-flavor Schwinger model in the presence of a chemical potential. Our ansatz uses a layered structure that can easily incorporate the relevant symmetries of the Hamiltoinan by simply restricting certain parameters in each layer. Moreover, we demonstrated that the ansatz lends itself to both both circuit-based and measurement-based quantum hardware.

Focusing on the case of three fermion flavors, we simulated the VQE using our ansatz classically for various parameter regimes, assuming a perfect quantum computer without any noise. This performance benchmark of the ansatz demonstrated that it can approximate the ground state of the model well, even in regimes where conventional MC methods suffer from the sign problem. Specifically, we can resolve the first-order phase transitions that are present in the model and reliably capture the phase structure with our ansatz circuit. Moreover, our results for different system sizes indicate that the  number of layers required to achieve a good performance scales only moderately with the number of lattice sites in the model.

To demonstrate the suitability of our ansatz for gate-based quantum hardware, we performed inference runs on IBM's superconducting quantum devices. To this end we used a set of parameters obtained from classically simulating the VQE, and prepared the resulting state on the quantum hardware to measure the energy and the particle number. To compensate for part of the hardware noise, we used ZNE to mitigate some of these effects. Despite ZNE not being able to fully mitigate the hardware noise, we were still able to reliably identify the different phases in the investigated area of the phase diagram via the particle numbers. Moreover, we were able to determine the critical points using the quantum hardware results. The resulting numerical values lie very close to the exact critical points, allowing us to determine them with good accuracy from the noisy hardware results.

In our proof-of-principle run on a quantum hardware we only used ZNE to mitigate hardware noise, yielding results are in good agreement with the theoretical expectation. Thus, carefully using more elaborate error mitigation methods such as Pauli twirling, readout error mitigation, measurement error mitigation and dynamical decoupling, we expect that our ansatz can be scaled up to larger system sizes on current quantum hardware. A systematic investigation of our ansatz for larger system sizes and performing a full VQE on quantum hardware will be done in future work. Moreover, while the effects of hardware noise and error mitigation can be straightforwardly studied in the circuit model, error mitigation for measurement-based quantum computers is a lot less explored. In the future, we also plan to investigate the potential of our ansatz for realistic, noisy measurement-based devices and to explore the possibility to mitigate errors on such quantum hardware.

\acknowledgments
S.K.\ acknowledges financial support from the Cyprus Research and Innovation Foundation under the project ``Quantum Computing for Lattice Gauge Theories (QC4LGT)'', contract no.\ EXCELLENCE/0421/0019.
This work is funded by the European Union’s Horizon Europe Framework Programme (HORIZON) under the ERA Chair scheme with grant agreement no.\ 101087126 and by the Deutsche Forschungsgemeinschaft (DFG, German Research Foundation) – Project-ID 429529648 – TRR 306 QuCoLiMa (“Quantum Cooperativity of Light and Matter’’).
This work is supported with funds from the Ministry of Science, Research and Culture of the State of Brandenburg within the Centre for Quantum Technologies and Applications (CQTA). 
\begin{center}
    \includegraphics[width = 0.08\textwidth]{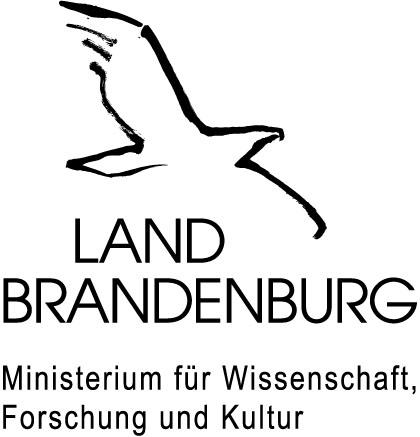}
\end{center}
\bibliography{papers}
\end{document}